\documentclass[10pt,twocolumn,twoside, romanappendices]{IEEEtran}

\usepackage{amsmath}
\usepackage{amsfonts}

\usepackage[dvips, final]{graphicx}
\usepackage{color}
\usepackage{caption}
\usepackage{subfigure}

\usepackage{latexsym}
\usepackage{pstricks}
\usepackage{pst-plot}
\usepackage{epsfig}
\usepackage{epstopdf}

\usepackage{algorithm}
\usepackage{algpseudocode}
\usepackage{amssymb}

\begin{document}
\title{Optimal Frame Transmission for Scalable Video with Hierarchical Prediction Structure}

\author{Saied Mehdian and Ben Liang$^*$
\thanks{The authors are with the Department of Electrical and Computer Engineering, University of Toronto, 10 King's College Road, Toronto, Ontario, Canada. Emails: \{smehdian,liang\}@comm.utoronto.ca.  This work has been supported in part by a grant from Bell Canada and the NSERC CRD program.}

}

\maketitle

\begin{abstract}
An optimal frame transmission scheme is presented for streaming scalable video over a link with limited capacity. The objective is to select a transmission sequence of frames and their transmission schedule such that the overall video quality is maximized. The problem is solved for two general classes of hierarchical prediction structures, which include as a special case the popular dyadic structure. Based on a new characterization of the interdependence among frames in terms of trees, structural properties of an optimal transmission schedule are derived.  These properties lead to the development of a jointly optimal frame selection and scheduling algorithm, which has computational complexity that is quadratic in the number of frames.  Simulation results show that the optimal scheme substantially outperforms three existing alternatives. 

\end{abstract}

\section{Introduction\label{S0}}

Video streaming is contributing to an ever increasing portion of the Internet traffic going through service providers.  Besides stringent demands on bandwidth and delay, it requires adaptation to heterogeneous access networks and devices, in order to achieve satisfactory viewing experience.  The efficient and adaptive transmission of video is paramount to service providers and users alike.

Toward this goal, one thrust of engineering effort is on efficient video coding. The H.264 Advanced Video Coding (AVC) standard, including its Scalable Video Coding (SVC) enhancements \cite{ITU-T2007}\cite{Schwarz2007}, is the most widely adopted coding scheme, used in a wide array of applications including Bluray and Youtube. It provides a means to encode a video stream into sub-streams of different quality, which may be selectively transmitted and decoded based on the available communication hardware and bandwidth.  However, the H.264 standard does not specify how to select or schedule video data for transmission.

The other engineering thrust is on adaptive transmission schemes for scalable video streaming.  It concerns the joint optimization of two procedures, first, the selection of a subset of the video frames for transmission, and second, scheduling the transmission of those frames before their display deadlines expire.  This is challenging mainly due to the \emph{Hierarchical Prediction Structure (HPS)} of the video codec \cite{Schwarz2007}.  Prediction is necessary to improve coding efficiency, but it creates complicated dependencies between frames in a video sequence.  Most existing studies on efficient transmission of scalable video \cite{Schierl2009}\nocite{schierl2011priority}\nocite{Zhang2010}\nocite{Du2010}\nocite{Stuhlmuller2000}\nocite{Xu2007}\nocite{Li2009}-\cite{Li2010a} address dependence by a simplified \emph{flow-based} model, in which a video layer can be decoded only if all lower layers are already decoded. In the more realistic \emph{frame-based} approach, dynamic frame dropping and retransmission have been studied in \cite{Feng1999}\nocite{Kritzner2004}\nocite{Liebl2005a}\nocite{Liebl2006}\nocite{Chou2006}\nocite{Badia2010}-\cite{Fu2010}, but these proposals are mostly based on heuristics that are challenging to analyze mathematically.

In this work, we present an analytical approach to optimize the frame transmission schedule of videos with hierarchical prediction, with joint consideration for the optimal frame selection given a limited link capacity.  The optimization objective is to minimize the loss of playback quality in the transmitted video sequence. To the best of our knowledge, this is the first work to provide a provably optimal polynomial solution to frame-by-frame lossy video transmission which accounts for hierarchical prediction.

Our main contributions are as follows.  \emph{First}, we formally characterize the inherent properties of an optimal frame transmission schedule in terms of the dependency structure arising from the HPS. This leads to the categorization of two general classes of videos, termed Sequential Isomorphically Ordered and Quasi Sequential Isomorphically Ordered, which include as special cases some commonly employed prediction schemes such as the hierarchical dyadic structure \cite{Schwarz2007}.  \emph{Second}, we develop efficient algorithms to compute the frame ordering rules in an optimal transmission schedule.  We show that they hold regardless of the subset of frames selected for transmission and, in particular, they govern a unique universal transmission sequence of which an optimal schedule is a subsequence.   \emph{Third}, based on the above structural observations, we propose dynamic programming solutions for jointly optimal frame selection and scheduling, which is shown to have only quadratic complexity.  Simulation with common test video traces show that the proposed method can substantially improve the quality of lossy video streaming over a link with limited capacity.

The rest of this paper is organized as follows. In Section \ref{S2}, the related works are presented. In Section \ref{S1}, we explain the system model and problem statement. In Section \ref{Pp}, we discuss the special properties of an optimal schedule.  In Sections \ref{OS} and \ref{exten}, optimal algorithms for scheduling in the two general classes of HPS are presented.  In Section \ref{S4}, simulation results with video traces are shown to demonstrate the performance gain by the proposed optimal solution.  Finally, Section \ref{S5} concludes the paper.

\section{Related Work\label{S2}}

Previous works related to this paper can be categorized into two groups: classical scheduling and scalable video streaming.

\subsection{Classical Deterministic Scheduling}

Classical scheduling problems related to our work are addressed under the theory of deterministic scheduling with delay constrained jobs \cite{Graham1979},\cite{Lawler1993},\cite{pinedo2012scheduling}.  The schedule may need to satisfy some precedence relation between jobs.  However, if a job is discarded, it has no effect on the other jobs \cite{Lawler1969}.  This is different from the dependency relation between video frames.  To the best of our knowledge, there is no optimal algorithm with polynomial complexity on scheduling with dependency.

\subsection{Scheduling for Scalable Video Streaming}

Many prior works do not consider the dependency relation in video transmission \cite{Li2005a}\nocite{Huang2008}\nocite{li2008content}\nocite{Liang2007}\nocite{Liang:TMM2008}\nocite{Dua2007}\nocite{dua2007downlink}\nocite{dua2010channel}-\cite{Kalman2004}. In all theses works the video content is transmitted in an earliest-deadline-first (EDF) fashion.  In \cite{li2008content}, a dropping scheme is proposed which does not take into account deadline contraints.  In particular, for each regular time interval, a constant size set of consecutive frames is considered for transmission and some of them are dropped to comply with channel rate limit, regardless of deadline constraints. In \cite{Li2005a}, \cite{Huang2008}, a dropping scheme  utilizing dynamic prorgramming is presented which dynamically adjusts the transmission power and therefore the rate to ensure selected packets meet their deadlines. Studies that do consider the dependency relation can be divided into two main groups: flow based and frame based.
 In the flow-based approach \cite{Schierl2009}\nocite{schierl2011priority}\nocite{Zhang2010}\nocite{Du2010}\nocite{Stuhlmuller2000}\nocite{Xu2007}\nocite{Li2009}-\cite{Li2010a}, the video is modelled as a set of inter-dependent data flows, each providing basic or enhanced playback quality. Flow-based dependency structure often is relevant only within the same display frame.  In this work, we consider the more complicated frame-based dependency structures.

In the frame-based approach, the video sequence is modelled as a set of data units, roughly corresponding to the display frames. Each data unit may depend on one or more data units for decoding.  In
\cite{Feng1999}\nocite{Kritzner2004}\nocite{Liebl2005a}-\cite{Liebl2006}, heuristics are proposed to drop data units under bad link conditions according to some pre-defined priority.  No analytical result is presented on how to set priority levels, and evaluation is performed through simulation only.  In comparison, our proposed scheduling algorithm is provably optimal.

More complex online frame-based transmission schemes for dependent frames in the lossy environment have been studied \cite{Chou2006,Badia2010,Fu2010,Jurca2007}.
These works target more ambitious problems than ours, since they need to account for the uncertainty in the system data, including factors such as multipath congestions and buffer overflows.  Optimal solutions are generally intractable, and only heuristic solutions are available.  In this work, we target a simpler offline scheduling problem that is suitable for streaming pre-recorded videos over a stable link and propose an efficient but jointly optimal frame selection and scheduling algorithm.

Furthermore, we provide an optimal scheme permitting the transmission 
of frames past their display deadline, for the benefit of displaying other frames. 
This is not available in any of the above works.
More detail are provided in Subsection \ref{qoullc1}.

\section{Hierarchical Prediction Structure and Problem Statement\label{S1}}

In this section, we first overview the Hierarchical Prediction Structure widely employed in various video codecs.
We then state the problem of video quality optimization under limited link capacity and discuss the decision space of the problem.

\subsection{Hierarchical Prediction Structure}

The video consists of a sequence of frames indexed in their display order by $1, 2,\ldots, N$.
The frames are classified into three groups: I-frames, P-frames, and B-frames. I-frames are intra-coded and do not depend on other frames, P-frames are inter-coded based on a preceding frame in the display order, and B-frames are inter-coded based on a preceding frame and a successive frame.  The set of P-frames and B-frames between two consecutive I-frames, plus the leading I-frame, is called a group of pictures (GOP).  The P-frames and B-frames of different GOPs are isolated from each other, but those within a GOP adhere to a specific dependency structure governed by the adopted prediction coding.  This corresponds to the general HPS of H.264 AVC, which enables temporal scalability \cite{Schwarz2007}.

The H.264 AVC standard is one of the most popular video compression techniques in the industry. For example, Blu-ray discs and players must support the H.264 AVC codec. In addition,
this standard is widely used by video-streaming service providers such as Vimeo, Youtube, and iTuneStore, and by various HDTV broadcasting standards. The SVC extension to
H.264 provides a richer set of methods to achieve scalability, but that is at the cost of significantly increased  implementation complexity \cite{Schwarz2007}\cite{Richardson2010}. A common
practical option is to operate SVC with only a single spatial or quality layer \cite{Schwarz2007}.

\begin{figure}[tbp]
   \begin{center}
    \epsfig{file=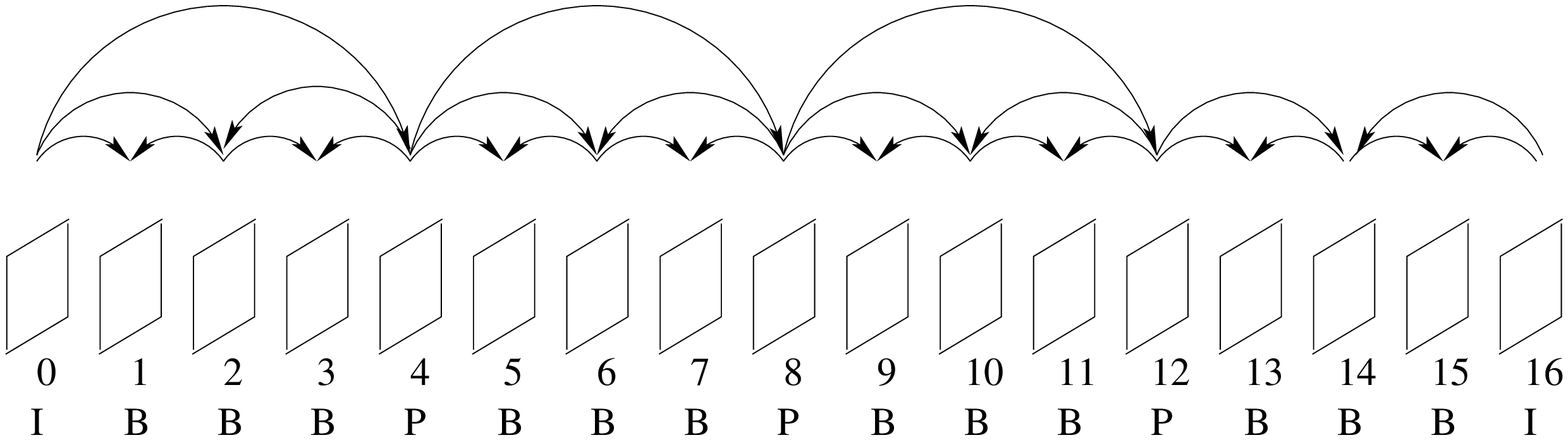, scale=0.35}
   \end{center}
\caption{Frames $(0,1, \ldots, 15)$ form a GOP with hierarchical dyadic structure $G16B3$.}
  \label{fig:fig1}
\vspace*{-5mm}
\end{figure}

In this work, we focus on common videos with \textit{temporal} scalability.
As an example, the most widely adopted HPS is the \emph{hierarchical dyadic structure}.  Following the notations of \cite{Seeling2011}, a hierarchical dyadic structure is denoted by $GnBm$ where $n$ is the size of each GOP and $m$ is the number of B-frames between consecutive I-frames or P-frames, with $m=2^{\omega}-1$ for some $\omega \in \mathbb{N}$.  Figure \ref{fig:fig1} illustrates a GOP with $G16B3$.

The dependency between frames is represented as a directed acyclic graph (DAG), where frames are nodes, and their dependencies are indicated by edges \cite{Chou2006}, as shown in Figure \ref{fig:fig1}.  If decoding frame $l$ requires frame $l'$ \emph{directly}, which is denoted by $l'\rightarrow l$, then a direct edge connects node $l'$ to node $l$. In this case,  $l'$ is called a \emph{parent} of $l$, and $l$ a \emph{child} of $l'$.  If the decoding of frame $l$ depends on frame $l'$, possibly through some intermediate nodes, then $l'$ is called an \emph{ancestor} of $l$, and $l$ a \emph{descendant} of $l'$. If two frames have no ancestral relation, they are called \emph{irrelevant} and denoted by $l \nLeftrightarrow l'$.  Throughout this paper, we use $G$ to denote the DAG corresponding to the video sequence under consideration, and the terms ``frame'' and ``node'' are used interchangeably depending on context.

We denote the sequence of GOPs as $\{GOP_i\}$ in their display order.
We also denote the sequence of I-frames as $\{I_i\}$ in their display order, so that $I_i$ corresponds to $GOP_i$. The descendants of $I_{i+1}$ can only be in $GOP_i$ and $GOP_{i+1}$, and except $I_{i+1}$, the frames in $GOP_i$ cannot depend on any frame in other GOPs.  Let ${\cal N}_i$ be the set of non-I-frames in $GOP_i$, and ${\cal D}_i$ be the set of frames in $GOP_i$ that are descendants of both $I_i$ and $I_{i+1}$.

\subsection{Quality Optimization under Limited Link Capacity \label{qoullc1}}

We consider the last hop of an access link to a video streaming user of a wired/wireless network. In this time-slotted scenario, the transmitter sends a pre-recorded video sequence to the receiver through a link with fixed capacity $C$ bits/timeslot.  We assume that a lower-layer protocol ensures the correct reception of any data transmitted at or below the link capacity.  Without loss of generality, we omit the propagation delay on the link, since otherwise we only need to shift the display time of all frames by a constant offset to accommodate it. Moreover, we note that access links commonly incur propagation delay on the order of micro seconds, which is negligible compared with the frame dispaly period of a video sequence.

Each video frame $l$ is associated with three parameters $(S_l, d_l, q_l)$, where $S_l$ is its size in bits, $d_l$ is its display deadline offset against some given transmission start time (i.e., time 0), and $q_l$ is its quality increment, the expected loss of video quality if the frame is not displayed \cite{Chou2006}.  For example, one possible measure of $q_l$ is the peak signal-to-noise ratio.
Denote the deadline of the last frame as $\widehat{T}$.

A \emph{transmission schedule} is a vector containing the transmission starting times of a sequence of video frames, sorted in ascending order. It indicates both the selected frames for transmission and the order of transmission.
In a transmission schedule, a frame is \emph{decodable} if and only if its parents are already decoded.  A frame is \emph{successful} if it becomes decodable and arrives at the receiver prior to its display deadline.

The reward function of a transmission schedule, $\mathcal{S}$, is defined as the sum of the quality increment of its successful frames, i.e.,
\begin{equation}
z(\mathcal{S}) = \sum_{l \in \mathcal{S}, \; l \text{ is successful}}{q_l}
\end{equation}
Our objective is to find a transmission schedule that maximizes the reward given a link capacity limit $C$.
This problem is called the \emph{Scheduling Problem} throughout the rest of this paper.  The resultant transmission schedule is called an \emph{optimal schedule}, and the resultant reward function is indicated with $z^*$.

\textbf{Transmission of Unsuccessful Frames.}
An optimal schedule may permit unsuccessful frames to be transmitted. Although these frames are not displayed, their importance arises from the fact that due to the dependency structure, they can help decode other frames. However, the transmission of an unsuccessful frame which has no successful descendant is useless. Therefore, in order to obtain an optimal schedule, we need to focus only on transmission schedules in which each unsuccessful frames have at least a successful descendant.

\textbf{Sufficiency of One-by-One Frame Transmission.}
The following theorem indicates that it suffices to consider only one-by-one frame transmission in the search for an optimal schedule.
\newtheorem{thm0001}{Theorem}
\begin{thm0001}
Every schedule can be transformed into a schedule with only one-by-one frame transmission, with at least the same successful frames and therefore, with at least the same reward function.
\label{thm0001}
\end{thm0001}
\begin{proof}
See Appendix \ref{AP1}.
\end{proof}

In this case, the number of timeslots required to transmit frame $l$ is given by $\Delta t_l=\frac{S_l}{C}$.  We assume the timeslot size is small enough such that $\Delta t_l$ are well approximated by integers.

\textbf{Optimal Transmission Sequence.}
With one-by-one frame transmission, there is no benefit in leaving a gap between the transmission of two consecutive frames, in terms of the reward function. Neither is there a penalty in leaving a gap, as longs as the gap does not lead to out-of-date transmissions. Therefore, in computing an optimal schedule, it suffices to consider only the \textit{transmission sequence}, i.e., the set of frames selected for transmission and their order of transmission. With an optimal transmission sequence, an optimal schedule can be determined by simply transmitting the frames back-to-back without any gap between them. Hence, throughout the rest of this paper, a ``transmission schedule" refers to only its ``transmission sequence" without the timing information.

\begin{table}[tbp]
  \caption{Table of Terminology}
  \label{Tbl1}

\centering
\begin{tabular}{|c|p{5.5 cm}|}

  \hline
  Term & Definition \\
 \hline
  $l \nLeftrightarrow l'$ & Frames $l$ and $l'$ are irrelevant \\
  \hline
  $l' \rightarrow l$ & Frame $l$ depends on frame $l'$ \\
  \hline
   $GOP_i$ & The $i$-th GOP \\
  \hline
   $I_i$ & The $i$-th I-frame \\
  \hline
   ${\cal N}_i$ & The set of non I-frames in $GOP_i$ \\
  \hline
   ${\cal D}_i$& The set of frames dependent on both $I_i$ and $I_{i+1}$ \\
  \hline
   $d_l$ & The deadline of frame $l$ \\
  \hline
   trasmission schedule & A vector containing the transmission starting times of a sequence of video frames, sorted in ascending order \\
  \hline
   decodable frame & A frame with none of its ancestors missing \\
  \hline
   successful frame & A decodable frame which arrives at the receiver prior to its deadline \\
  \hline
   $z({\cal S})$ & Reward function of transmission schedule ${\cal S}$\\
  \hline
   transmission sequence & The set of frames selected for transmission and their order \\
  \hline
   $T(x)$ & The subtree rooted at node $x$\\
  \hline
   $min\_dln(T)$ & Minimum display deadline in rooted tree $T$ \\
  \hline
   $max\_dln(T)$ & Maximum display deadline in rooted tree $T$ \\
  \hline
   $c_i(x)$ & The $i$-th child of node $x$ \\

  \hline

\end{tabular}
\end{table}

The decision space of this problem includes all permutations of all subsets of frames in the video sequence. Therefore, the complexity of exhaustive search would be prohibitive.  In the next three sections, we first present some inherent properties of an optimal transmission schedule, which will then be used in Sections \ref{OS} and \ref{exten} to develop polynomial solutions to the scheduling problem.
A partial list of the terminology is provided in Table \ref{Tbl1}.

\section{Properties of an Optimal Schedule}\label{Pp}

In this section, we first describe a transformation of the dependency DAG of a video into rooted trees that indicate decodability. Based on this representation, we then present two general classes of HPS that are of interest, and give some important properties of optimal scheduling common to both classes and essential to the solutions presented in Sections \ref{OS} and \ref{exten}.

\subsection{Modified Breadth First Search Trees}

We adopt a version of the Breadth First Search (BFS) algorithm \cite{Cormen2001} on the video DAG $G$, which we call Modified Breadth First Search (MBFS).  It takes $G$ and a node $s$ as input and outputs an \emph{MBFS tree} rooted at $s$. The main difference with BFS is the following: at each node, instead of picking all unvisited children of that node, only \emph{decodable} unvisited children are picked, where a node is decodable if and only if all of its ancestors have been visited. Moreover, in constructing the MBFS tree, the decodable unvisited children are sorted in ascending order of their deadlines.  Such ordering is important to the concept of isomorphically ordered trees presented later.
The MBFS algorithm is formally given in Appendix \ref{alg1}.

\begin{figure}[tbp]
   \begin{center}
    \epsfig{file=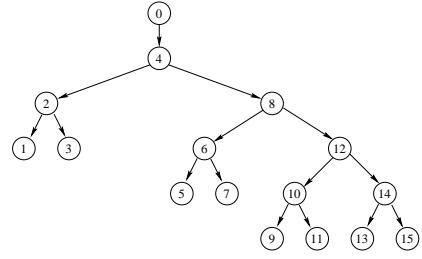, scale= 0.3}
   \end{center}

\caption{MBFS on DAG in Figure \ref{fig:fig1} with node 16 removed.}
  \label{fig:fig2}
  \end{figure}

We run MBFS on $G$ and each I-frame in the display order, creating an \emph{MBFS forest}, whose components are MBFS trees rooted at the I-frames, each tree corresponding to a GOP.  The complexity of this procedure is $O(N)$, since each MBFS tree creation has constant complexity given a fixed GOP size. Figure \ref{fig:fig2} illustrates the result of MBFS on the DAG in Figure \ref{fig:fig1} with node 16 removed. 
The main benefit of MBFS is to represent the dependency structure of frames in the format of trees.

We emphasize that in terms of the MBFS forest, the children of each node, in standard graph theoretic terminology for trees, is a subset of the children of that node in terms of the DAG $G$, as defined in Section \ref{S1}.  This is because being a child in the MBFS forest carries the additional requirement of being decodable.  Furthermore, the set of descendants of an I-frame $I_i$ in terms of the DAG can be partitioned into two subsets in $GOP_{i-1}$ and $GOP_i$, where only the latter subset is the descendants of $I_i$ in terms of the MBFS forest.  In the rest of this paper, when we refer to the parent-child and ancestor-descendant relations between frames, they are in terms of the DAG by default.  When the relations are in terms of the MBFS forest, such exceptions will be clearly stated unless they are obvious from the context.

However, note that the set of descendants of a non-I-frame remains the same in terms of either the MBFS forest or the DAG.
Hence, if we focus our attention within a single GOP, then any frame has the same set of descendants in terms of either the MBFS forest or the DAG.  In that case we only need to distinguish the reference terms of children but not those of descendants.

\subsection{SIO and Quasi-SIO Classes}

Based on the MBFS forest, we next give two important definitions that characterize the HPSs of interest.

\newtheorem{def1}{Definition}
\begin{def1}
The DAG representing the HPS of a video is called \emph{sequential} if the ancestors of each node lie only on the path between the node and the root of the MBFS tree that contains the node.
\end{def1}

Thus, if a DAG is sequential, then decoding a node only requires the availability of nodes residing on the path between the node and the root of the MBFS tree that contains the node.  Clearly, if a video sequence consists of only I-frames and P-frames, then its DAG is sequential, since each node only has a single parent.  The situation is more complicated when there are B-frames.  From Figure \ref{fig:fig2}, it is easy to see that the DAG in Figure \ref{fig:fig1} is sequential if and only if edges $16\rightarrow 14$ and $16\rightarrow 15$ are removed.

In the MBFS forest, let $T(x)$ denote the subtree rooted at node $x$, and $min\_dln(T(x))$ and $max\_dln(T(x))$ be the minimum and maximum display deadlines among the nodes in $T(x)$, respectively.  Let $c_i(x)$ be the $i$th child of $x$.

\newtheorem{def2}[def1]{Definition}
\begin{def2}
An \emph{isomorphically ordered tree} is a rooted tree such that, if node $x$ has $k$ children, they can be re-ordered (and re-indexed) such that
\begin{equation}
 max\_dln(T(c_i(x))) \leq min\_dln(T(c_{i+1}(x))) , 1 \leq i \leq k-1 ~.
\end{equation}
\end{def2}

The above definition resembles that of B-trees \cite{Cormen2001}, but there is no requirement for the tree being balanced. An example of isomorphically ordered trees are binary search trees.  A more general example is shown in Figure \ref{fig:fig2}.  We confine our discussions to HPSs whose MBFS forest contain only isomorphically ordered trees.
The following lemma will be used in deriving the optimal schedules.
\newtheorem{thm8}{Lemma}
\begin{thm8}
In an MBFS forest with isomorphically ordered trees, consider two irrelevant nodes $x$ and $y$. If $d_x<d_y$, then $max\_dln(T(x)) <  min\_dln(T(y))$.
\label{thm8}
\end{thm8}

\begin{proof}
See Appendix \ref{AP10}.
\end{proof}

From the two definitions above, we say that an HPS is \emph{Sequential and Isomorphically Ordered} (SIO) if its DAG is sequential and its MBFS forest contains only isomorphically ordered trees.  For example, the HPS shown in Figure \ref{fig:fig1} is SIO if  edges $16\rightarrow 14$ and $16\rightarrow 15$ are removed.   Furthermore, to extend the concept of SIO to more general HPSs that contain B-frames that depend on an I-frame from the next GOP, we say that an HPS is \emph{Quasi Sequential and Isomorphically Ordered} (quasi-SIO) if it would become SIO after such dependency was removed.  For example, the HPS shown in Figure \ref{fig:fig1} is quasi-SIO.

In this work, we focus only on the SIO and quasi-SIO classes of HPSs, since many common HPSs in practice belong to either class.  For example, the zero-delay structure \cite{Schwarz2007} is SIO and the classical prediction structure \cite{Seeling2011} used as default in H.264/AVC is quasi-SIO.  For a proof that the widely adopted hierarchical dyadic structure is quasi-SIO, see Appendix \ref{S7}.

\subsection{Canonical Form of Optimal Transmission Sequence \label{cFOT1}}

Theorem \ref{Prp1} below suggests that it suffices to find an optimal transmission sequence that satisfies some special properties.  In particular, for the SIO and quasi-SIO classes, we say that a transmission sequence that preserves the two properties in Theorem \ref{Prp1} is in the \emph{canonical form}.

\newtheorem{Prop1}[thm0001]{Theorem}
\begin{Prop1}
In the SIO and quasi-SIO classes, any transmission sequence $\cal{S}$ can be re-ordered into a transmission sequence  $\cal{S'}$ with at least the same set of successful frames (i.e., $z(\cal{S})$ $ \leq z(\cal{S'})$) and with the following properties:
\begin{enumerate}
  \item  The ancestors of a frame are scheduled prior to that frame.
  \item  Consider two irrelevant frames $l_i$ and $l_j$ with $d_{l_i}<d_{l_j}$.  If they are neighbors in the transmission sequence, then $l_i$ is scheduled before $l_j$.  If $l_i$ is scheduled after $l_j$, then not all of the frames scheduled between $l_j$ and $l_i$ are irrelevant with respect to $l_j$.
\end{enumerate}
\label{Prp1}
\end{Prop1}
\begin{proof}
See Appendix \ref{AP2}.
\end{proof}

Theorem \ref{Prp1} will be used extensively in the rest of this paper.  Furthermore,
the lemmas below will also have important usage in Sections \ref{OS} and \ref{exten}.

\newtheorem{thm001}[thm8]{Lemma}
\begin{thm001}
Under the properties in Theorem \ref{Prp1}, consider the frames $f_1,o_1,\ldots,o_k$ such that $f_1 \in GOP_i$, none of $o_1,\ldots, o_k$ belongs to $GOP_i$, and they are all irrelevant with respect to $f_1$.

\begin{enumerate}
 \item If they are scheduled in the order of $f_1,o_1,\ldots, o_k$, then

\begin{equation}
 d_{f_1} < d_{o_r}, 1\leq r \leq k
\end{equation}

and all frames $\{o_r\}_{r=1}^{k}$ belong to GOPs with indices greater than $i$.

\item If they are scheduled in the order of $o_1,\ldots, o_k, f_1$, then

\begin{equation}
  d_{f_1} > d_{o_r}, 1\leq r \leq k
\end{equation}

and all frames $\{o_r\}_{r=1}^{k}$ belong to GOPs with indices less than $i$.
\end{enumerate}

\label{lm1}
\end{thm001}

\begin{proof}
See Appendix \ref{AP3}.
\end{proof}

\newtheorem{thm7}[thm8]{Lemma}
\begin{thm7}
    Under the properties in Theorem \ref{Prp1}, consider some frames $f_1,o_1,\ldots,o_k $ in an MBFS tree, which corresponds to a GOP.  Suppose all of $o_1,\ldots, o_k$ are irrelevant with respect to $f_1$.

\begin{enumerate}
 \item If they are scheduled in the order of $f_1,o_1,\ldots, o_k$, then

\begin{equation}
 d_{f_1} < d_{o_r}, 1\leq r \leq k.
\end{equation}

\item If they are scheduled in the order of $o_k,\ldots, o_1, f_1$, then

\begin{equation}
  d_{f_1} > d_{o_r},  1\leq r \leq k.
\end{equation}

\end{enumerate}

\label{thm7}
\end{thm7}

\begin{proof}
See Appendix \ref{AP5}.
\end{proof}

\section{Optimal Schedule for the SIO Class\label{OS}}

This section provides an algorithm to solve the Scheduling Problem for the SIO class.  We first show that each transmission sequence has a unique canonical-form transmission sequence as defined in Subsection \ref{cFOT1}. Noting that such a sequence is a subsequence of some general pattern that we term the \emph{universal sequence}, we then select an optimal schedule from the universal sequence through dynamic programming.

\subsection{SIO Universal Sequence\label{sq1}}

First, we discuss how to find a canonical-form transmission sequence $\cal W$ given an arbitrary transmission sequence.
The lemma below indicates that the frames of each GOP must be transmitted together in $\cal W$. Moreover, due to the second property of Theorem \ref{Prp1}, GOPs with earlier deadlines must be transmitted first.

\newtheorem{lma0}[thm8]{Lemma}
\begin{lma0}
In an SIO canonical-form transmission sequence, for any $i$, no frame other than the frames of $GOP_i$ can be scheduled between the frames of $GOP_i$.
\label{lma0}
\end{lma0}

\begin{proof}
See Appendix \ref{AP12}.
\end{proof}

For the frames inside a GOP, the properties of isomorphically ordered MBFS trees in Section \ref{Pp} can be used to determine the order of transmissions.  Lemma \ref{thm5} indicates that all subtrees in an MBFS tree must be scheduled back-to-back in order to preserve Theorem \ref{Prp1}.

\newtheorem{thm5}[thm8]{Lemma}
\begin{thm5}
In an SIO canonical-form schedule, any frame scheduled between any two frames that are both in a subtree $T$ must itself be in $T$.
\label{thm5}
\end{thm5}
\begin{proof}
See Appendix \ref{AP6}.
\end{proof}

In $\cal W$, consider an arbitrary node $x$ and the subtrees $T(c_1(x)),\ldots,T(c_{n}(x))$, where $c_1(x), \ldots, c_{n}(x)$ are the children of $x$ in the MBFS forest sorted in ascending order of display deadlines. The frame $x$ is sent first according to the first property of Theorem \ref{Prp1}. Then, by Lemma \ref{thm5}, the second property of Theorem \ref{Prp1}, and isomorphic order, the frames in each $T(c_{i}(x))$ are sent together in the order of $c_i(x)$.   Therefore, to determine the order of transmissions for each GOP in $\cal W$, we should perform a generalized \emph{pre-order tree walk} \cite{Cormen2001} on each tree of the MBFS forest, where the children of each node are visited in ascending order of display deadlines.  This procedure uniquely determines $\cal W$ and has a complexity of $O(N)$. Therefore, a canonical form transmission sequence is uniquely determined only by the set of selected frames for transmission.

Next, consider the sequence of all frames of the original video. We call the canonical-form transmission sequence of this sequence the \emph{SIO universal sequence}.  Note that since we may hypothetically increase the link capacity until the entire video is successfully schedulable, such a transmission sequence always exists.

\newtheorem{thm10}[thm0001]{Theorem}
\begin{thm10}
For the SIO class of HPSs, the canonical-form of any transmission sequence is a subsequence of the SIO universal sequence.
\label{thm10}
\end{thm10}

\begin{proof}
See Appendix \ref{AP13}.
\end{proof}

Theorem \ref{thm10} suggests that, given a link capacity limit, a transmission sequence that is a subsequence of the SIO universal sequence and which maximizes the playback quality is an optimal transmission sequence.
This will be used in the next subsection to compute an optimal schedule.

\subsection{Computation of Optimal Schedule}  \label{sec:SIO_compute}

The following dynamic programming approach solves the Scheduling Problem for the SIO class.
First, generate the SIO universal sequence and index its frames as $1,2,\ldots,N$.

Then, define function $h(j,t)$ as the maximum reward function, if frames $\{j,\dots,N\}$ are to be scheduled in the time interval $[t,\widehat{T}]$ assuming all their parents with indices in the range from $1$ to $j-1$ (if any) are available.
From the system model in Section \ref{S1}, we have
\begin{align}
z^{*}&=h(1,0)
\end{align}
and the boundary conditions
\begin{align}
h(N+1,t)&=0, \qquad \forall t \in \mathbb{Z}\label{f1}\\
h(j,t)&=0, \qquad t>\widehat{T} .\label{f2}
\end{align}

Furthermore, $h$ adheres to the following recursive equations, corresponding to the possible actions at time $t$:
\begin{equation}
\label{rec1}
h(j,t)= \max \begin{cases}
h(j,t+1)\\
q_j+h(j+1,t+\Delta t_j), \quad d_j-t \geq \Delta t_j  \\
h(j+1,t+\Delta t_j), \quad d_j-t < \Delta t_j  \\
h(\min\{k :  k>j, k \nLeftrightarrow j \},t).
\end{cases}
\end{equation}
In the above, $h(j,t)$ is set to be the best outcome among the four possible actions. The first term corresponds to the case where the optimal schedule for $h(j,t+1)$ is also optimal for $h(j,t)$, so no action is needed at $t$. The second term corresponds to starting to transmit frame $j$ at time $t$, and the transmission is successful, which is ensured by the condition $d_j-t \geq \Delta t_j$. In this case, we gain $q_j$ toward the objective along with the reward for all future frames. 
Similarly, the third term refers to starting to transmit frame $j$ at time $t$ and the transmission is not successful as indicated by the condition $d_j-t < \Delta t_j$ (although the frame's transmission may  potentially help its descendants to achieve successful transmission). The forth term corresponds to dropping frame $j$ and moving to inspect the next frame in the universal sequence that does not depend on $j$.

Since $h$ only needs to be computed for  $ 1\leq j \leq N, 0 \leq t \leq \widehat{T}$, dynamic programming requires $O(N \widehat{T})$ processing time and $O(N \widehat{T})$ memory to determine $h(1,0)$ and extract the optimal policy.  Furthermore, since $\widehat{T}$ is linear in $N$ in all practical video codecs with a constant frame rate, we have an overall complexity of $O(N^2)$.

\section{Optimal Schedule for the Quasi-SIO Class\label{exten}}

In this section, an optimal transmission sequence for the more general quasi-SIO class is presented, as an extension of the solution in Section \ref{OS}.

\subsection{Quasi-SIO Universal Sequence}

Again, we will first show that all transmission sequences have unique canonical-form transmission sequences, and then they will be shown to be subsequences of a quasi-SIO universal sequence.

Let $\cal V$ be a canonical-form transmission sequence of an arbitrary transmission sequence in the quasi-SIO class.
From the first property of Theorem \ref{Prp1}, $I_i$ will be scheduled prior to the frames in ${\cal N}_i$ since $I_i$ is the ancestor of all of them, and similarly $I_{i+1}$ is scheduled prior to the frames in ${\cal D}_i$.

Consider the following lemmas:

\newtheorem{lma1}[thm8]{Lemma}
\begin{lma1}
In $\cal V$, if $f_1$ and $f_2$ are two frames in ${\cal N}_i$ such that no other frame in ${\cal N}_i \cup \{I_{i+1}\}$ is scheduled between them, then no frame can be scheduled between them.
\label{lma1}
\end{lma1}

\begin{proof}
The proof is similar to that of Lemma \ref{lma0}.
\end{proof}

\newtheorem{lma2}[thm8]{Lemma}
\begin{lma2}
In $\cal V$, if $I_{i+1}$ is scheduled after $f_1 \in {\cal N}_i$ such that no other frame in ${\cal N}_i$ is scheduled between them, then no frame can be scheduled between them.
\label{lma2}
\end{lma2}

\begin{proof}
See Appendix \ref{AP14}.
\end{proof}

\newtheorem{lma3}[thm8]{Lemma}
\begin{lma3}
In $\cal V$, if $I_{i+1}$ is scheduled before $f_2 \in {\cal N}_i$ such that no other frame in ${\cal N}_i$ is scheduled between them, then no frame can be scheduled between them, and $f_2$ must be a descendant of $I_{i+1}$.
\label{lma3}
\end{lma3}

\begin{proof}
See Appendix \ref{AP15}.
\end{proof}

\newtheorem{lma4}[thm8]{Lemma}
\begin{lma4}
In $\cal V$, if $I_{i+1}$ is scheduled before $f \in {\cal N}_{i+1}$ such that no other frame in ${\cal N}_{i+1}$ is scheduled between them, then only frames in $GOP_i$ and frame $I_{i+2}$ can be scheduled between them.
\label{lma4}
\end{lma4}

\begin{proof}
See Appendix \ref{AP16}.
\end{proof}

Lemmas \ref{lma1}, \ref{lma2}, and \ref{lma3} jointly indicate that the frames of ${\cal M}_i={\cal N}_i \cup \{I_{i+1}\}$ must be sent together. Moreover, Lemma \ref{lma4} indicates that ${\cal M}_{i+1}$ must be scheduled immediately after ${\cal M}_i$, for all $i$.

Next, we determine the transmission order of frames inside each ${\cal M}_i$.  Due to the first property of Theorem \ref{Prp1}, each I-frame $I_{i+1}$  is scheduled prior to ${\cal D}_i$ in $\cal V$. With respect to $I_{i+1}$, the frames of $GOP_i$ can be divided into two groups: $\mathcal{A}_i$, the frames scheduled prior to $I_{i+1}$, and $\mathcal{B}_i$, the frames scheduled after $I_{i+1}$. Clearly all frames of ${\cal D}_i$ are in $\mathcal{B}_i$.  Since the frames in ${\cal M}_i$ must be sent together,  the frames in $\mathcal{A}_i$ are sent together, and the same holds for $\mathcal{B}_i$.
The definition of quasi-SIO indicates that the frames of $\mathcal{A}_i$ can be scheduled by Lemma \ref{thm5} using the subgraph of $G$ which comprises the nodes of $\mathcal{A}_i$.  Moreover, for $\mathcal{B}_i$ since $I_{i+1}$ has already been scheduled, the frames of $\mathcal{B}_i$ can be scheduled similarly based on the isomorphic ordering property, the second property of Theorem \ref{Prp1}, and Lemma \ref{thm5}.  Hence, all subtrees in $\mathcal{B}_i$ are transmitted as contiguous blocks in the ascending order of the display deadlines of their roots.

\begin{figure}[tbp]
 \begin{center}
  \scalebox{0.3}{\input{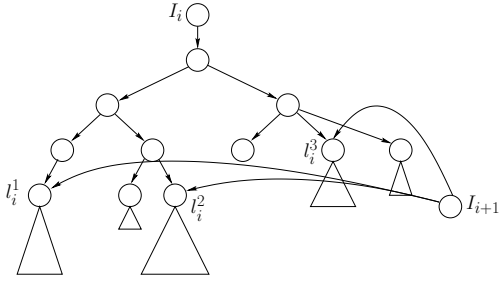}}
\caption{Critical nodes in the $GOP_i$ MBFS tree of the quasi-SIO class, plus edges from $I_{i+1}$. Each triangle indicates a subtree rooted at the node located at its top.}
\label{fig:fig14}
 \end{center}
 \vspace*{-4mm}
\end{figure}

It remains to determine the relative position of the frames in ${\cal N}_i-{\cal D}_i$ with respect to $I_{i+1}$.
Let a critical node be defined as $f$, $f \in {\cal D}_i$, such that in the path between $f$ and the root of the MBFS tree containing $f$, no other frame in ${\cal D}_i$ appears. An example of this is depicted in Figure \ref{fig:fig14}, where $l^1_{i},l^2_{i}$ and $l^3_{i}$ are critical nodes. It is easy to see that a critical node in $GOP_i $ is a child of $I_{i+1}$, but a child of $I_{i+1}$ in $GOP_i$ is not necessarily a critical node.  Let $\Gamma_i$ be the set of all critical nodes.   It is clear from Figure \ref{fig:fig14} that the set of nodes in the subtrees rooted at the critical nodes of $GOP_i$ is equivalent to ${\cal D}_i$. Furthermore, due to the definition, the members of $\Gamma_i$ are pair-wise irrelevant. Let $\zeta_i$ be the member of $\Gamma_i$ with the smallest display deadline.

\newtheorem{lma6}[thm8]{Lemma}
\begin{lma6}
 In $\cal V$, frame $\zeta_i$ must be scheduled immediately after $I_{i+1}$.

 \label{lma6}
\end{lma6}

\begin{proof}
See Appendix \ref{AP11}.
\end{proof}

Furthermore, we have the following result.
\newtheorem{lma5}[thm8]{Lemma}
\begin{lma5}
 In $\cal V$, consider frame $f \in {\cal N}_i-{\cal D}_i$ that is irrelevant with respect to $\zeta_i$.
\begin{itemize}
  \item If $d_f < d_{\zeta_i}$, $f$ is scheduled before $I_{i+1}$.
  \item If $d_f > d_{\zeta_i}$, $f$ is scheduled after $I_{i+1}$.
\end{itemize}

 \label{lma5}
\end{lma5}

\begin{proof}
See Appendix \ref{AP8}.
\end{proof}

The above observations indicate that for any quasi-SIO transmission sequence, if  we first ignore the backward prediction edges from $I_{i+1}$ to B-frames in $GOP_i$, for all $i$, and create the SIO canonical form transmission sequence, then the quasi-SIO canonical form schedule can be obtained by simply moving $I_{i+1}$, for all $i$, in the SIO canonical form transmission sequence backward so that it is positioned immediately prior to the earliest frame of ${\cal D}_i$ in the transmission order. Furthermore, $\cal V$ is uniquely determined given any transmission sequence. Hence, we define the \emph{quasi-SIO universal sequence} similarly to the SIO universal sequence in the previous section.

Then, similar to Secion \ref{OS}, we have the following theorem, which will be exploited to derive the optimal schedule given a link capacity limit.  
\newtheorem{thm11}[thm0001]{Theorem}
\begin{thm11}
For the quasi-SIO class of HPSs, the canonical-form of any transmission sequence is a subsequence of the quasi-SIO universal sequence.
\label{thm11}
\end{thm11}

\begin{proof}
See Appendix \ref{AP17}.
\end{proof}

\subsection{Computation of Optimal Schedule}

First, generate the quasi-SIO universal sequence and index the frames by $1, 2, \ldots, N$.

Then, define function $g(j,t,s)$, similarly to $h(j,t)$, as the maximum sum quality of successful frames, if frames $\{j,\dots,N\}$ are to be scheduled in the time interval $[t,\widehat{T}]$ assuming all their parents with indices in the range from $1$ to $j-1$ (if any) are available.  The additional parameter $s$ is an integer in $\{0,1,2,3\}$, whose binary representation specifies the status of the two nearest I-frames in the quasi-SIO universal sequence that precede $j$, i.e., $s = (x_1(j), x_2(j))$, where $x_2(j) = 1$ if the nearest preceding I-frame is selected for transmission and 0 otherwise, and $x_1(j)$ is similarly defined concerning the second nearest preceding I-frame.

Similarly to Subsection \ref{sec:SIO_compute}, we have
\begin{align}
z^{*}&=g(1,0,0),
\end{align}
with boundary conditions
\begin{align}
g(N+1,t,s)&=0, \qquad \forall t \in \mathbb{Z}, \forall s \label{f10}\\
g(j,t,s)&=0, \qquad t>\widehat{T}. \label{f20}
\end{align}
The recursive equations for $g(j,t,s)$ are given below, with special consideration for the parameter $s$.

1) If $j$ is not an I-frame, let the index of the GOP that $j$ belongs to be $i_j$.  Three cases are considered:

1a) Suppose $j \in \mathcal{A}_{i_j}$.  This is the case where $j$ does not depend on $I_{i_j+1}$ and is located prior to $I_{i_j+1}$ in the quasi-SIO universal sequence.  If $I_{i_j}$, the nearest preceding I-frame, is transmitted, then the outcome is similar to Section \ref{OS}; otherwise, $j$ should not be selected.  Hence, for $s=1,3$, we have
\begin{equation}
g(j,t,s)= \max \begin{cases}
g(j,t+1,s)\\
q_j+g(j+1,t+\Delta t_j,s), \quad d_j-t \geq \Delta t_j \\
g(j+1,t+\Delta t_j,s), \quad d_j-t < \Delta t_j \\
g(\min\{k :  k>j, k \nLeftrightarrow j \},t,s)
\end{cases}
\label{eq:g_trans}
\end{equation}
and for $s=0,2$, we have
\begin{equation}
g(j,t,s)= \max \begin{cases}
g(j,t+1,s)\\
g(\min\{k :  k>j, k \nLeftrightarrow j \},t,s)
\end{cases}
\label{eq:g_notrans}
\end{equation}

1b) Suppose $j \in {\cal D}_{i_j}$.  This is the case where $j$ depends on both $I_{i_j}$ and $I_{i_j+1}$ and is located behind both in the quasi-SIO universal sequence.  If both I-frames are transmitted, then the outcome is similar to Section \ref{OS}; otherwise, $j$ should not be selected.  Hence, for for $s=3$, we use \eqref{eq:g_trans}, and for $s=0,1,2$, we use \eqref{eq:g_notrans}.

1c) Suppose $j \in \mathcal{B}_{i_j} - {\cal D}_{i_j}$.  This is the case where $j$ does not depend on $I_{i_j+1}$ but is located behind $I_{i_j+1}$ in the quasi-SIO universal sequence.  If $I_{i_j}$, the \emph{second} nearest preceding I-frame, is transmitted, then the outcome is similar to Section \ref{OS}; otherwise, $j$ should not be selected.  Hence, for $s=2,3$, we use \eqref{eq:g_trans}, and for $s=0,1$, we use \eqref{eq:g_notrans}.

2) If $j$ is an I-frame, then besides scheduling the frame similar to Subsection \ref{sec:SIO_compute}, we also need to update $s$. It is easy to verify that, if the frame is dropped, then we update to $(2s) \text{ mod } 4$; if the frame is transmitted, then we update to $(2s+1) \text{ mod } 4$.  Hence,

\begin{equation}
g(j,t,s)= \max \begin{cases}
g(j,t+1,s)\\
q_j + g(j+1, t+\Delta t_j,(2s+1) \text{ mod } 4), \\
   \hspace*{10ex} d_j-t \geq \Delta t_j  \\
g(j+1, t+\Delta t_j,(2s+1) \text{ mod } 4), \\
   \hspace*{10ex} d_j-t < \Delta t_j  \\
g(\tilde{I}_j,t,(2s) \text{ mod } 4) ,
\end{cases}
\end{equation}
where $\tilde{I}_j$ is the next I-frame after $j$, and we update the two bits of $s$ depending on the scheduling outcome.

In the above dynamic programming formulation, state $s$ needs to take only four values due to the fact that the frames of a GOP only depend on at most two I-frames. Therefore, we again have complexity $O(N\widehat{T})=O(N^2)$.

\section{Experiments with Video Traces\label{S4}}

Even though the proposed schedule has been proven optimal, we are interested in its numeric performance when applied to real video traces.

\subsection{Methodology}
The proposed scheme is simulated in Matlab, using H.264 video traces provided by \cite{Seeling2011}.  We present here results for the ``NBC News ($352\times 288$)'' and ``SONY ($1920\times 1080$)'' traces, representing faster moving and slower moving scenes, respectively.   We take the first 305 frames of each video.  
The details of the video traces are provided in Table \ref{Tbl2}.  
The luminance of a frame is used as its quality measure, so the objective function averaged over the number of frames represents the average Y-PSNR at the receiver \cite{Stuhlmuller2000}.

\begin{table}
  \caption{Video Trace Specifications}
  \label{Tbl2}
  
\centering 
\begin{tabular}{|c|c|c|}

  \hline
  Sequence name & Sony$\_$1080 & NBC News \\
  \hline 
  Resolution & 1920$\times$1080 & 352$\times$288 \\
  \hline  
  FPS & 30 & 30 \\
  \hline 
  Encoder & JSVM (9.15) & JSVM (9.19.14) \\
  \hline 
  Encoding Type & High (Level 5) & High (Level 2.1) \\ 
  \hline 
  GoP Pattern & G16B15 & G16B15 \\
  \hline
  Quantization Parameters (I,P,B) & 28,N/A,30 & 29,N/A,37 \\
  \hline 
  Layer & 4 & 4 \\
  \hline
\end{tabular}
\end{table}

We compare the optimal solution with three suboptimal alternatives. The first is a best-effort earliest-deadline-first (EDF) algorithm, in which the frames are scheduled in their display order, and the frames that cannot be successful are dropped by the transmitter.
EDF is the traditional scheme used for video streaming \cite{schierl2011priority}. The second is a decoding order earliest-deadline-first (DOEDF) algorithm, which is similar to the first algorithm except the frames are transmitted in decoding order. The third is a priority-based earliest-deadline-first (PBEDF) algorithm, which combines frame prioritization  \cite{Feng1999}\cite{Kritzner2004} with EDF.  In PBEDF, the set of frames are partitioned into blocks of size $M$.  In each block, the I-frames are scheduled first in EDF order, and the B-frames are scheduled last in EDF order.  For fair comparison, in the results below we always use an optimal $M$ obtained by exhaustive search.
Note that for a constant value of $M$, the PBEDF algorithm, similar to the EDF and DOEDF algorithms,  has a complexity of $O(N)$. It should be clear that 
all three alternatives are suboptimal because their transmission decisions do not take in account the next frames in the sequences. In our experimental study, we aim to quantify the 
performance gain of the optimal policy in order to justify the added computational complexity.

\subsection{Experimental Results}

\begin{figure}[tbp]
  \centering
  \subfigure[initial delay=0.1 sec]{\includegraphics[scale=0.3]{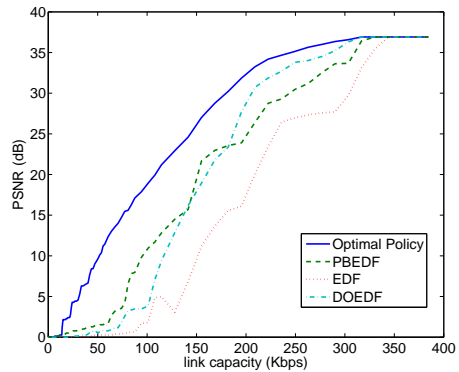}}
  \subfigure[initial delay=1 sec]{\includegraphics[scale=0.3]{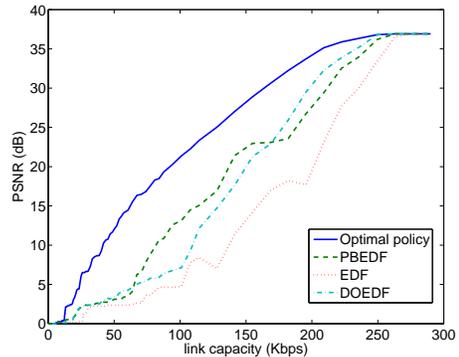}}
  \subfigure[initial delay=5 sec]{\includegraphics[scale=0.3]{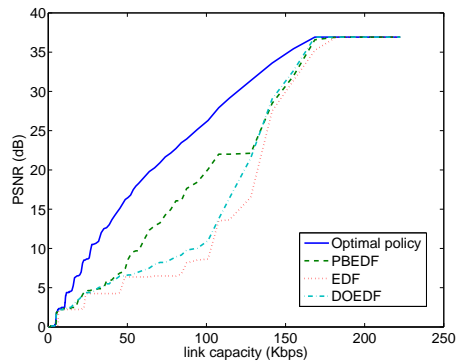}}
  \caption{Y-PSNR per video frame vs.~link capacity for video NBC News ($352\times 288$).}
  \label{fig:figsing_NBC}
\end{figure}
\begin{figure}[tbp]
  \centering
  \subfigure[initial delay=0.1 sec]{\includegraphics[scale=0.3]{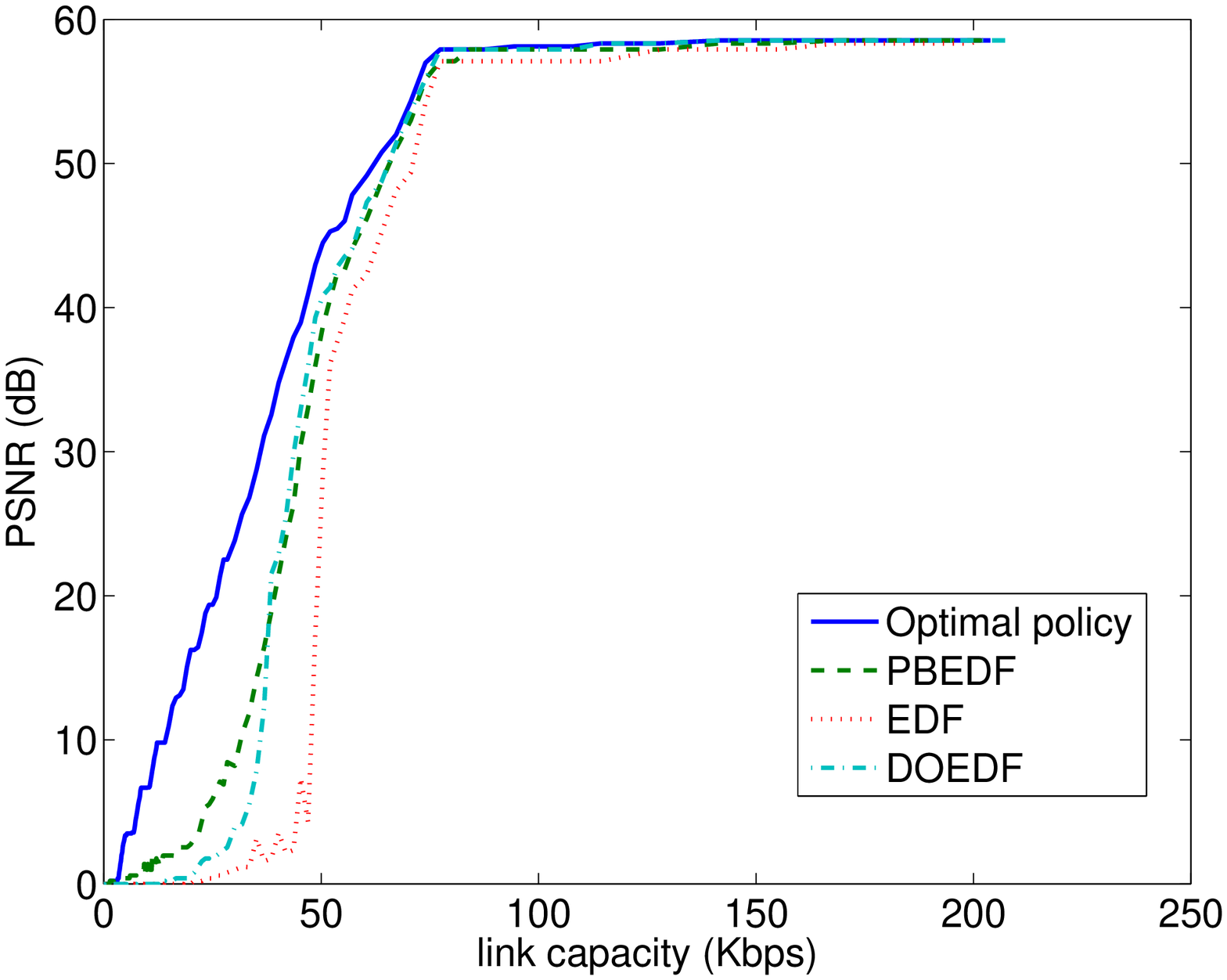}}
  \subfigure[initial delay=1 sec]{\includegraphics[scale=0.3]{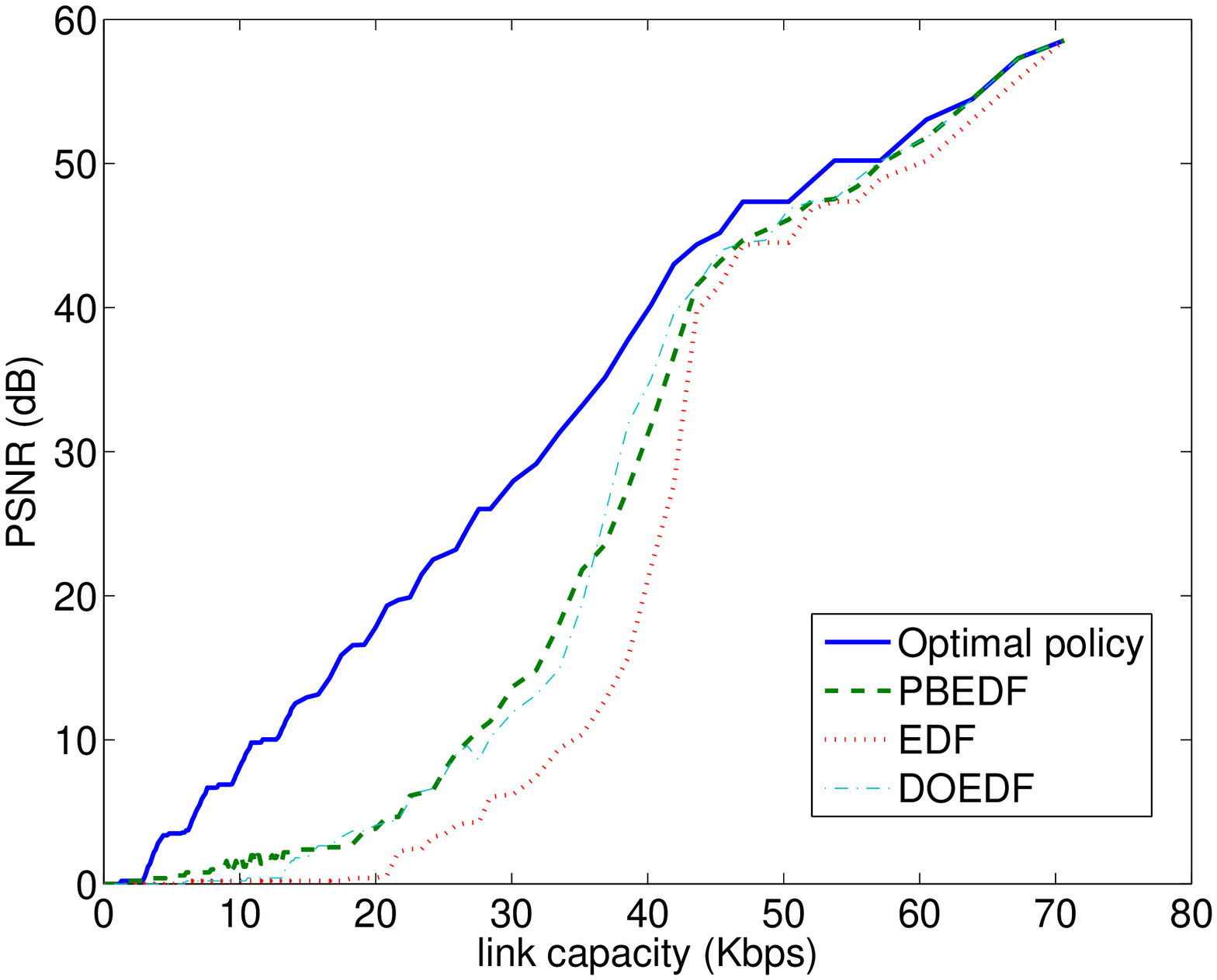}}
  \subfigure[initial delay=5 sec]{\includegraphics[scale=0.3]{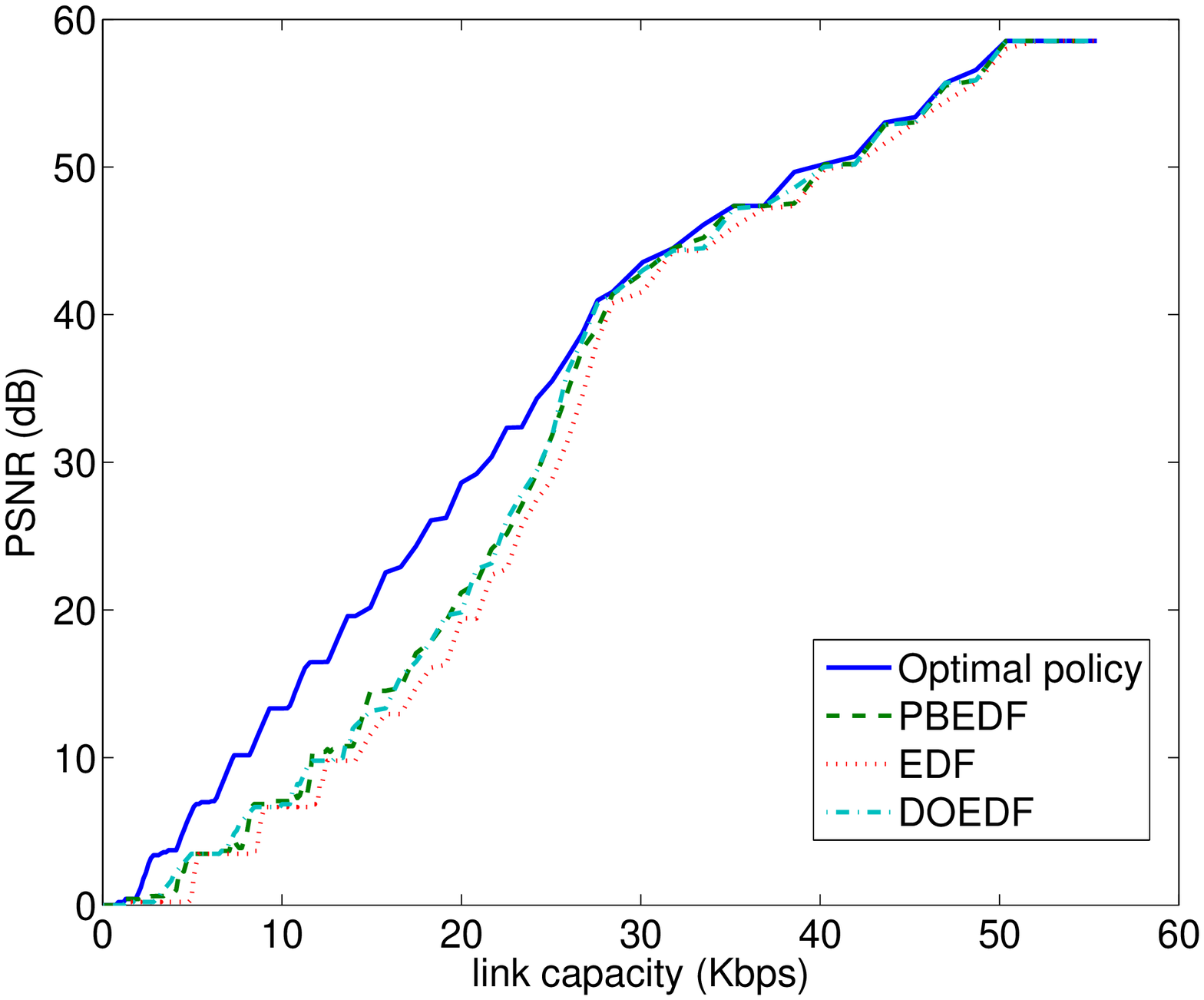}}
  \caption{Y-PSNR per video frame vs.~link capacity for video SONY ($1920\times 1080$).}
  \label{fig:figsing_SONY}
\end{figure}

Video playback quality is dictated by the link rate and the initial playback delay, which is defined as the time the receiver waits before video playback while receiving data.   In Figures \ref{fig:figsing_NBC} and \ref{fig:figsing_SONY}, we consider different initial delays of 0.1, 1, and 5 seconds, and for each initial delay, we study all values of link capacity $C$ up to the point when all algorithms give lossless performance.  Note that since the video trace ``SONY'' has very low motion, its data size is relatively smaller, so that high PSNR can be achieved even for its HD resolution.

We observe that, for a wide range of parameter settings, the optimal schedule substantially outperforms the sub-optimal alternatives.  The exception is only under extremely relaxed environments, e.g., when the initial delay is large or when the link capacity is high.  In practical video streaming, where the users are inpatient, and the network bandwidth limited, the benefit of the optimal schedule is apparent.

In addition, under the optimal schedule, the Y-PSNR is monotonically increasing in the link capacity as expected.  This is not the case for EDF, DOEDF and PBEDF, which results from the dependence structure.  In particular, a slightly higher link capacity may drive EDF, DOEDF and PBEDF to over zealously transmit a frame, which in turn reduces the time left to transmit an ancestor that is later in the display order, possibly rendering the ancestor unsuccessful.  Hence, with these sub-optimal schemes, increasing the link capacity is not always beneficial.

\section{Conclusion and Discussion\label{S5}}

A frame scheduling scheme has been developed for optimal video streaming over a link with limited capacity.  For two general classes of HPSs, it is shown that an optimal transmission sequence can be found as a subsequence of some optimal universal sequence in term of frame indices.  Efficient dynamic programming solutions are proposed to identify the optimal schedules with polynomial complexity.  Experiments with video traces show that the optimal schedules can substantially improve the playback quality over sub-optimal alternatives.

The proposed scheme is optimal for a wide range of practical codecs including MPEG4, H.264 AVC and single spatial layer implementation of SVC.
The problem of jointly optimization of multiple spatial layers is outside the scope of this paper. However, we note that the proposed scheme can be used 
to design efficient transmission schemes for multiple spatial layers, by applying it to each layer seperately.

\appendices

\section{Proof of theorem \ref{thm0001} \label{AP1}}
There are two possible scenarios where the frames are not transmitted one-by-one in some part of a schedule.   First, the transmission of a frame may be interrupted by the transmission of parts of other frames (preemptive transmission).  Second, parts of multiple frames are transmitted simultaneously.  We present in the following how to transform the schedule in both scenarios to one-by-one frame transmission without incurring new delay in the transmission end time of any frame.

\begin{figure}[tbp]
 \begin{center}
  \scalebox{0.31}{\input{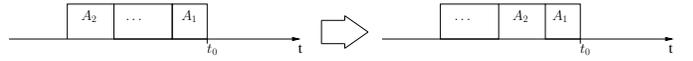}}
\caption{Transforming a preemptive schedule to a non-preemptive schedule.}
\label{fig:fig8}
 \end{center}
\end{figure}

In the first scenario, as illustrated in Figure \ref{fig:fig8}, supposed $A_1$ and $A_2$ are consecutive fragments of a frame, and some other frames (or parts of frames) are scheduled between them.  Clearly, swapping the transmission of $A_2$ and that of the interrupting frames, as depicted in the right side of Figure \ref{fig:fig8}, will not cause new delay in the transmission end time of any frame. Therefore, all preemptive transmissions can be removed in this way.

\begin{figure}[tbp]
 \begin{center}
  \scalebox{0.31}{\input{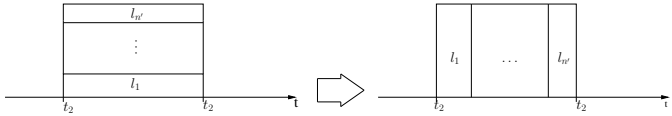}}
\caption{Transforming simultaneous transmissions to nonsimultaneous transmissions}
\label{fig:fig9}
 \end{center}
\end{figure}

In the second scenario, we first consider the case where frames $l_1,\ldots, l_{n'}$ exactly overlap each other, within the transmission interval $[t_1,t_2]$.  As depicted in Figure \ref{fig:fig9}, if these frames are sent one-by-one under the same link capacity, then clearly the amount of data transmitted in this interval will not change and all the frames can be transmitted before time $t_2$. Next, consider the case where the frames partially overlap each other.  In this case, time can be divided into subintervals, such that in each of them, the frame segments completely overlap each other.   Then the same transformation above can be carried out in each subinterval, to produce a schedule without simultaneous transmissions but with preemptive transmissions.  This is then reduced to the first scenario.

\section{The Modifed Breadth First Search Algorithm \label{alg1}}

\begin{algorithm}{}
\caption{Modified Breadth\- First Search MBFS(G,$\,$s)}
\begin{algorithmic}[1]
    \For{each node $u \in V[G]-\{s\}$}
        \State $color[u]\leftarrow WHITE$
	\State $d[u]\leftarrow \infty$
	\State $\pi[u]\leftarrow NULL$
    \EndFor

    \State $colors[s]\leftarrow GRAY$
    \State $d[s]\leftarrow 0$
    \State $Q\leftarrow \emptyset$
    \State ENQUEUE($Q,s$)

  \While {$Q\neq \emptyset$}
  \State $u\leftarrow$ DEQUEUE(Q)
  \State $Adj[u] \gets$ sequence of children of node $u$
  \State sort $Adj[u]$ in ascending order of deadlines.
  \For{each $v \in Adj[u]$(in ascending order)}
  \If {($color[v]=WHITE$) and ($\forall i , i\rightarrow v \Rightarrow color[i]=GRAY$)}
  \State $color[v]\leftarrow GRAY$
  \State $d[v]\leftarrow d[u]+1$
  \State $\pi[v]\leftarrow u$
  \State ENQUEUE($Q,v$)
  \EndIf
  \EndFor
  \EndWhile

  \State return
\end{algorithmic}
\end{algorithm}

\section{Proof of Lemma \ref{thm8} \label{AP10}}

\newtheorem{thm9}[thm8]{Lemma}
\begin{thm9}
Let $x$ and $y$ be two irrelevant nodes in an isomorphically ordered tree. A unique node $w$ and unique indices $i$ and $j$, $i\neq j$, exist such that $x$ belongs to $T(c_i(w))$ and $y$ belongs to $T(c_j(w))$.
 \label{thm9}
\end{thm9}

\begin{proof}
Let $u_1,\ldots,u_{m(x)}$ and $v_1,\ldots,v_{m(y)}$ correspond to the paths from the root of the tree to $x$ and $y$ respectively. Clearly $u_1 = v_1$ is the root, but the two sequences diverge later. Therefore, there is a unique index $\sigma$ such that $u_{\sigma}=v_{\sigma}$ but $u_{\sigma+1} \neq v_{\sigma+1}$.  Clearly, $u_{\sigma+1}$ and $v_{\sigma+1}$ are children of $u_{\sigma}$ in the tree. In addition, $x$ resides in $T(u_{\sigma+1})$ while $y$ resides in $T(v_{\sigma+1})$.
\end{proof}

Now back to the proof of Lemma \ref{thm8}, if $x$ and $y$ do not belong to the same GOP, then the lemma clearly holds. Now suppose they belong to the same GOP.
Based on Lemma \ref{thm9}, there exist a node $w$ and indices $i$ and $j$ such that $x$ belongs to $T(c_i(w))$ and $y$ belongs to $T(c_j(w))$. If $i<j$, then the definition of isomorphically ordered trees gives
$max\_dln(T(c_i(w))) \leq min\_dln(T(c_j(w)))$, but since none of the playback deadlines are equal, we have $max\_dln(T(x)) < min\_dln(T(y))$. If $i>j$, it can be shown similarly $min\_dln(T(x)) > max\_dln(T(y))$ which is not possible due to $d_x<d_y$. As a result, $i<j$ and $max\_dln(T(x)) < min\_dln(T(y))$.  

\section{Hierarchical Dyadic Structure\label{S7}}

We first present a mathematical characterization of the hierarchical dyadic structure and then show that it belongs to the quasi-SIO class.

In the hierarchical dyadic structure $GnBm$, the GOP size $n$ is an integer power of 2, and $m$ is the number of B-frames between consecutive non-B-frames, with $m=2^{\omega}-1$ for some $\omega \in \mathbb{N}$.  Each GOP contains one leading I-frame and $\frac{n}{m+1} - 1$ P-frames.  Each P-frame depends on the previous I-frame/P-frame in the display order. The dependency structure among the B-frames between two consecutive non-B-frames is described by $\text{Dyadic-build}(i,j)$ in Algorithm \ref{alg2}, where $j-i$ is an integer power of 2.  Note that in this algorithm, the frames of the video sequence are indexed in the display order as integers.  

\begin{algorithm}{}
\caption{$\text{Dyadic-build}(i,\,j)$}
\label{alg2}
\begin{algorithmic}[1]
  \If {$|i-j| \leq 1$ or $\log_2 |i-j| \notin \mathbb{N}$}
      \State return
  \EndIf
    \State $i_0 \gets \frac{i+j}{2}$
    \State $i_0$ depends on $i,j$
    \State Dyadic-build$(i,\,i_0)$
    \State Dyadic-build$(i_0,\,j)$
\end{algorithmic}
\end{algorithm}

As an example, Figure \ref{fig:fig1} shows the hierarchical dyadic structure for a GOP with $G16B3$.
A sequence of frames $i, \ldots, j$ is called a \emph{complete sequence} if, except $i$ and $j$, none of the frames in the sequence have parents outside this sequence.  Thus, for a complete sequence, if the frames $i$ and $j$ are available, all frames in the sequence can be decoded.  An example of a complete sequence is a GOP.  It can easily be shown through induction that, in the context of the hierarchical dyadic structure, the size of any complete sequence is $2^{\omega'}+1$, for some $\omega' \in \mathbb{N}$.

The hierarchical dyadic structure does not belong to the SIO class.  This is because there is always a B-frame that is the descendant of two consecutive I-frames, which violates the sequential property. In Figure \ref{fig:fig1}, frame $14$ is an example of such a frame that violates the sequential property.

However, in what follows, we show that the hierarchical dyadic structure belongs to the quasi-SIO class.  This is accomplished by demonstrating that a modification on the hierarchical dyadic structure, which removes the backward DAG edges emanating from each I-frame to its children in the preceding GOP, results in a structure belonging to the SIO class. In this modified hierarchical dyadic structure, no B-frame depends on a succeeding I-frame.  For instance, this means in Figure \ref{fig:fig1}, frames $14$ and $15$ no longer depend on frame $16$. It is worth mentioning that this modification is used as an approximation in \cite{Badia2010}. 
However, in our work, we show that an optimal schedule can be obtained for the hierarchical dyadic structure without modification, since it is a special case of the quasi-SIO class.

\newtheorem{thm2}[thm0001]{Theorem}
\begin{thm2}
The DAG of a modified hierarchical dyadic structure is sequential.
\label{them2}
\end{thm2}

\begin{proof}
The MBFS tree corresponding to each GOP is always rooted at the GOP's leading I-frame. If all B-frames are removed, only some subtrees of the MBFS forest are removed, and the resultant DAG consisting of only I-frames and P-frames is clearly sequential.  Therefore, it suffices to only prove that the sequential property is preserved when the MBFS algorithm is executed on B-frames between consecutive non-B-frames.

We use mathematical induction. In particular, since the number of B-frames between two consecutive non-B-frames is $2^{\omega}-1$, for some $\omega \in \mathbb{N}$, induction is carried out on $\omega$.   In the following, we address two cases based on whether the latter non-B-frame is 1) an I-frame or 2) a P-frame.

\begin{figure}[tbp]
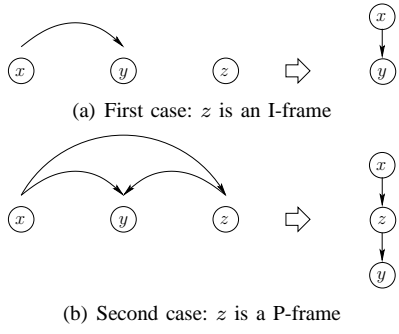

 \centering
  \subfigure[First case: $z$ is an I-frame]{\scalebox{0.4}{\input{fig10.pstex_t}}\label{fig:fig10}}
  \subfigure[Second case: $z$ is a P-frame]{\scalebox{0.4}{\input{fig15.pstex_t}}\label{fig:fig15}}
\caption{Hierarchichal dyadic structure with $\omega=1$ used in the proof of Theorem \ref{them2}. $x$ and $z$ are non-B-frames, and $y$ is a B-frame.}

\end{figure}

For the basis step, consider $\omega=1$.  This situation is depicted in Figures \ref{fig:fig10} and \ref{fig:fig15} for cases 1) and 2) respectively.  Note that in case 1), since frame $z$ is an I-frame, the edge from $z$ to $y$ is removed in the modified hierarchical dyadic structure.   In both cases, the MBFS tree clearly satisfies the sequential property.

For the induction step, suppose the statement holds for $\omega=k$.
This is shown in Figures \ref{fig:fig11} and \ref{fig:fig16} for cases 1) and 2) respectively.  In both cases, the number of frames between the two end frames is $2^k-1$, and the resulting tree satisfies the sequential property. The triangles in both figures indicate the subtree rooted at node $y$.  We consider $\omega=k+1$ for the two cases separately.

\begin{figure}[tbp]
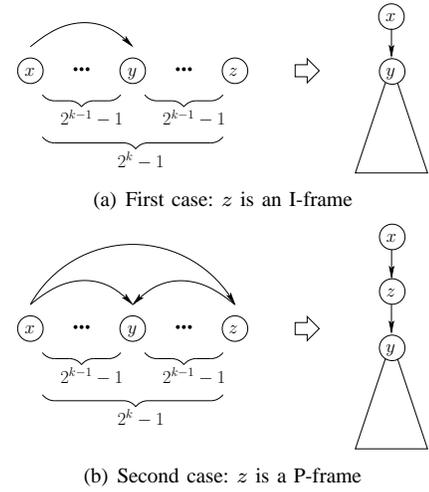

 \centering
  \subfigure[First case: $z$ is an I-frame]{\scalebox{0.4}{\input{fig11.pstex_t}}\label{fig:fig11}}
  \subfigure[Second case: $z$ is a P-frame]{\scalebox{0.4}{\input{fig16.pstex_t}}\label{fig:fig16}}
  \caption{Hierarchichal dyadic structure with $\omega=k$ used in the proof of Theorem \ref{them2}. $x$ and $z$ are non-B-frames, and all frames between them are B-frames.}

\end{figure}

\begin{figure}[tbp]
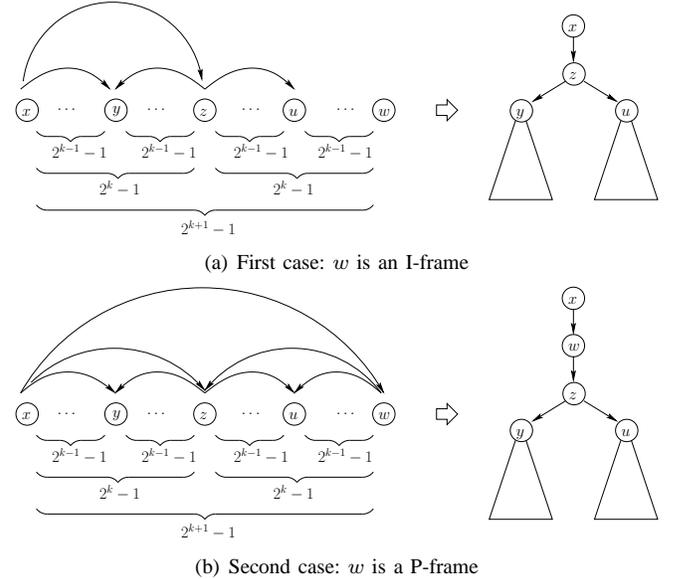

 \centering
  \subfigure[First case: $w$ is an I-frame]{\scalebox{0.35}{\input{fig12.pstex_t}}\label{fig:fig12}}
  \subfigure[Second case: $w$ is a P-frame]{\scalebox{0.35}{\input{fig17.pstex_t}}\label{fig:fig17}}
\caption{Hierarchichal dyadic structure with $\omega=k+1$ used in the proof of Theorem \ref{them2}. $x$ and $w$ are non-B-frames, and all frames between them are B-frames.}

\end{figure}

\emph{Case 1)}:

Figure \ref{fig:fig12} depicts this situation. There are $2^{k+1}-1$ nodes between $x$ and $w$. Node $z$ is located halfway between them. In addition, node $y$ is halfway between $x$ and $z$, and node $u$ is halfway between $z$ and $w$.  The dependency structure among these indicated nodes is displayed, and the rest of the dependency structure, which is not shown, is determined by the modified hierarchical dyadic structure.

The MBFS algorithm first visits $x$, then to $z$, $y$, and finally $u$.
It can be observed that the dependency structure for frames between $x$ and $z$ is the same as Figure \ref{fig:fig16}, and the dependency structure for frames between $z$ and $w$ is the same as Figure \ref{fig:fig11}.  Hence, the triangles in Figure \ref{fig:fig12} refer to the same triangles as in Figures \ref{fig:fig11} and \ref{fig:fig16}.

For nodes $x$, $y$, $z$, and $u$, it is easy to see that the sequential property holds. Within the subtree rooted at $y$, due to the inductive hypothesis in Figure \ref{fig:fig16}, the sequential property is satisfied.  Now consider the additional ancestors $x$ and $z$.  All paths from $x$ to nodes in the subtree rooted at $y$ pass through $z$ and $y$. Therefore for all nodes in the subtree rooted at $y$, the sequential property is preserved. Similarly, the same can be shown for the subtree rooted at $u$.

\emph{Case 2)}:

Figure \ref{fig:fig17} depicts this situation. Similar to the previous case, again there are $2^{k+1}-1$ nodes between nodes $x$ and $w$. Node $z$ is located halfway between them. Node $y$ is halfway between $x$ and $z$, and node $u$ is halfway between $z$ and $w$.  The dependency structure among these indicated nodes is displayed, and the rest of dependency structure, which is not shown, is determined by the modified hierarchical dyadic structure.

The MBFS algorithm first visits $x$.  Different from the previous case, however, it then visits $w$, before moving to $z$, $y$, and finally $u$.
When it visits $z$, nodes $x$ and $w$ have already been visited, so the remaining two branches of the dependancy structure, from $z$ to $y$ and from $z$ to $u$, each resembles that of \ref{fig:fig11}. In particular, $z$ in Figure \ref{fig:fig17} plays the role of $x$ in Figure \ref{fig:fig11}, while $y$ and $u$ in Figure \ref{fig:fig17} plays the role of $y$ in Figure \ref{fig:fig11}.

Similar to the previous case, for nodes $x$, $w$, $z$, $y$, and $u$, it is easy to see that the sequential property holds. Within the subtree rooted at $y$, due to the inductive hypothesis in Figure \ref{fig:fig16}, the sequential property is satisfied. Now consider the additional ancestors $x$, $w$, and $z$.  All paths from $x$ to nodes in the subtree rooted at $y$ pass through $w$, $z$, and $y$. Therefore for all nodes in the subtree rooted at $y$, the sequential property is preserved. Similarly, the same holds for the subtree rooted at $u$.

\end{proof}

\newtheorem{thm3}[thm0001]{Theorem}
\begin{thm3}
The MBFS trees under the hierarchical dyadic structure (modified or not) are Binary Search Trees (BSTs) with respect to the display deadlines.
\label{thm3}
\end{thm3}

\begin{proof}
Consider a hierarchical dyadic GOP with $GnBm$. To show that its corresponding MBFS tree is a BST, we need to verify the following three properties for every node in a BST:  the node has at most two children, the subtree rooted at its left child contains only nodes with deadlines earlier than that of the node, and the subtree rooted at its right child contains only nodes with deadlines later than that of the node.

Since in the creation of each MBFS tree, the dependencies on the succeeding I-frame does not apply, in what follows, we only need to consider a modified version of Algorithm \ref{alg2} that omits such dependencies.  Assume the video sequence is sorted and indexed in display order by integers.  We next consider each node in the DAG $G$ and show that when the MBFS algorithm visits the node, the BST properties are satisfied.  This is carried out for each type of nodes in the following.
\begin{itemize}
\item I-frame:  The I-frame is always the root of its own MBFS tree.  If $m<n-1$, then there exists at least one P-frame in the GOP.  After the MBFS algorithm visits this I-frame, among all of its children, only the first P-frame in display order is decodable, since the other children of the I-frame are also descendants of the first P-frame.   Hence, the I-frame has only a single child with respect to the MBFS tree, so the BST properties are satisfied.

    ~~If $m=n-1$, then the B-frame in the mid point of the GOP depends only on this I-frame in the modified structure, so it acts as a P-frame.  Similarly, this B-frame is the only child of the I-frame with respect to the MBFS tree, so the BST properties are satisfied.

\item P-frame:  Let the considered P-frame be $i_0$. The preceding and succeeding non-B-frames are the frames labelled ${i_0}-(m+1)$ and ${i_0}+(m+1)$ respectively. All children of $i_0$ are between these two frames.

    ~~Note that the subsequence $i_0-(m+1),\ldots,i_0$ is a complete sequence.  Among these frames, after the MBFS algorithm visits $i_0$, the only decodable child of $i_0$ is $\frac{(i_0-(m+1))+i_0}{2}$, since the other children of $i_0$ are all descendants of $\frac{(i_0-(m+1))+i_0}{2}$.
    The subsequence $i_0,\ldots,i_0+(m+1)$ is also a complete sequence, but we need to consider two cases depending on whether frame $i_0+(m+1)$ is an I-frame or a P-frame.  If $i_0+(m+1)$ is an I-frame, then after the MBFS algorithm visits $i_0$, the only decodable child of $i_0$ among $i_0, \ldots, i_0+(m+1)$ is $\frac{i_0+(i_0+(m+1))}{2}$, since the other children all depend on $\frac{i_0+(i_0+(m+1))}{2}$. If $i_0+(m+1)$ is a P-frame, then after the MBFS algorithm visits $i_0$, the only decodable child of $i_0$ among $i_0, \ldots, i_0+(m+1)$ is $i_0+(m+1)$, since the other children all depend on $i_0+(m+1)$.  In both cases, $i_0$ has only one decodable child among $i_0,\ldots, i_0+(m+1)$.

    ~~As a result, $i_0$ has exact two children in the MBFS tree. Moreover, the sequences $i_0-(m+1),\ldots,i_0-1$ and $i_0+1,\ldots, i_0+(m+1)$ are completely irrelevant, and all of the former have deadlines earlier than $d_{i_0}$ and all of the latter have deadlines later than $d_{i_0}$. Therefore, all BST properties are satisfied.

\item B-frame:  Let the considered B-frame be $i_0$.  This is similar to the previous case, except that instead of frames between $i_0-(m+1)$ and $i_0+(m+1)$, we should consider frames $i_0-(m_0+1), \ldots, i_0+(m_0+1)$ for some $m_0$ such that $m_0<\frac{m+1}{2}$ and $i_0-(m_0+1), \ldots, i_0+(m_0+1)$ is the largest complete sequence centered at $i_0$. We similarly see that, after the MBFS algorithm visits $i_0$, the only decodable children of $i_0$ are $\frac{(i_0-(m_0+1))+i_0}{2}$ and $\frac{i_0+(i_0+(m_0+1))}{2}$.  The other properties of BST are also similarly satisfied.
\end{itemize}
\end{proof}

The BSTs are a special case of isomorphically ordered trees.  Combining this with Theorem \ref{them2}, we see that the modified hierarchical dyadic structure belongs to the SIO class, and hence, the hierarchical dyadic structure belongs to the quasi-SIO class.

\section{Proof of Theorem \ref{Prp1} \label{AP2}}

Given a transmission sequence that does not meet either or both the two properties, it will be modified to meet those properties in two steps.

\begin{figure}[tbp]
 \begin{center}
  \scalebox{0.3}{\input{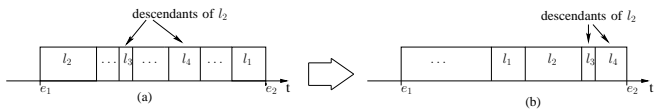}}
\caption{(a) A schedule in which $l_2$ is placed prior to $l_1$ but $l_1 \rightarrow l_2$. (b) Transformation to meet the first property.}
\label{fig:fig4}
 \end{center}
 \vspace*{-3mm}
\end{figure}

\emph{First Step:}  With the assumption that the first property does not hold, there must be frames $l_1$ and $l_2$ such that frame $l_1$ is the ancestor of frame $l_2$ but $l_2$ is scheduled before $l_1$.
An example of this is shown in Figure \ref{fig:fig4}(a). Let $K_{l_2}=\{\text{descendants of } l_2 \text{ scheduled after } l_2 \text{ and before } l_1\}$. Also let the index of the timeslot in which the transmission of $l_2$ starts be $e_1$ and the index of the timeslot in which the transmission of $l_1$ ends be $e_2$. If any frame in the set $K_{l_2}$ is successful, then since the transmission of frame $l_1$ has to terminate prior to the deadlines of its successful descendants, we have:

\begin{equation}
\quad e_2 \leq d_l, \forall l \in K_{l_2}\cup \{l_2\}, \text{ $l$ is successful}
\label{ty1}
\end{equation}

Now consider the frames in the interval $[e_1,e_2]$. They can be divided into three sets: $\{l_1\}$, $A_1=K_{l_2}\cup \{l_2\}$, and $A_2=\{\text{all frames in interval } [e_1,e_2]\}-(A_1 \cup \{l_1\})$.
Now starting from timeslot $e_1$ first schedule $A_2$, then $l_1$, and then $A_1$, such that the order of frames in $A_2$ in the original sequence is preserved and also the same for $A_1$. An example of this operation is shown in Figure \ref{fig:fig4}(b).

Any previously successful frame in the time interval $[e_1,e_2]$ remains successful. In particular, since all frames in $A_2$ either move backward in time or do not change position, the previously successful ones remain successful. Moreover, due to inequality \eqref{ty1} all previously successful frames in $A_1$  remain successful as well. Furthermore,  no new violation  of the first property is created.
Since this operation always repairs at least one pair that violates the first property, and the number of violations is finite, repetition of the operation will end in an solution which satisfies the first property.

\emph{Second Step:}
Let $U$ be the set of frames scheduled between $l_j$ and $l_i$.  Suppose all of them are irrelevant with respect to $l_j$.
Let $t_0$ be the timeslot when the transmission of $l_j$ starts and $t_1$ be the timeslot when the transmission of $l_i$ ends. If $l_i$ is successful, then $t_1 \leq d_{l_i} < d_{l_j}$.  

Else, if $l_i$ is not successful, then it has to have a successful descendant, namely $f_i$. Let $t_2$ be the timeslot when the transmission of $f_i$ ends. Since $f_i$ is successful, $t_2 \leq d_{f_i}$. Furthermore, from the \textit{First step}, we have $t_1 < t_2$. We next consider two cases based on whether $l_i$ is an I-frame.
If $l_i$ is not an I-frame, then $f_i$ belongs to the same tree as $l_i$ in the MBFS forest. Hence, Lemma \ref{thm8} indicates $d_{f_i} < d_{l_j}$.
If $l_i$ is an I-frame, let $n_{l_i}$ be the index of the GOP that $l_i$ belongs to. Then since $d_{l_i}<d_{l_j}$, the frame $l_j$ has to belong to a GOP with an index higher than $n_{l_i}$. But, the children of $l_i$ are only in $GOP_{n_{l_i}-1}$ and $GOP_{n_{l_i}}$. As a result, $d_{f_i} < d_{l_j}$. Hence, in both cases, we have $t_1 < t_2 < d_{l_j}$.

\begin{figure}[tbp]
 \begin{center}
  \scalebox{0.3}{\input{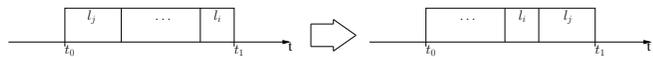}}
  \caption{Transformation corresponding to the second property in Theorem \ref{Prp1}.}
  \label{fig:fig5}
 \end{center}
 \vspace*{-6mm}
\end{figure}

Now, we can re-order the frames starting at $t_0$ as follows: $U$, then $l_i$, and then $l_j$. This clearly will not harm $l_i$ and $l_j$. In addition, the frames of $U$ are moved backward in time, so none of them is harmed either. Furthermore, none of the frames outside of $[t_0,t_1]$ are affected. This operation is illustrated in Figure \ref{fig:fig5}.  A special case of this is when $U$ is empty, i.e., when $l_j$ and $l_i$ are neighbors.

\vspace*{-3mm}

\section{Proof of lemmas \ref{lm1}-\ref{thm5}}

\subsection{Proof of Lemma \ref{lm1} \label{AP3}}

We use mathematical induction to prove the first part.

For the basis step, consider $k=1$.  Since $o_1$ and $f_1$ are from different GOPs and they are irrelevant, due to the second property of Theorem \ref{Prp1}, $d_{f_1} < d_{o_1}$ and $o_1$ belongs to a GOP with an index greater than $i$.

For the induction step, suppose the statement holds for $k=m$.  That is, $d_{f_1} < d_{o_m}$ and $o_m$ belongs to some $GOP_j$ with $j>i$. The frame $o_{m+1}$ either is a descendant of $o_m$ or is irrelevant with respect to $o_m$.

If $o_{m+1}$ is a descendant of $o_m$, then it belongs to $GOP_j$ or $GOP_{j-1}$.  In the former case, we have $j>i$ and hence $d_{f_1} < d_{o_{m+1}}$.  In the latter case, since $o_{m+1}$ and $f_1$ do not belong to the same GOP, $j-1 \neq i$.  Combining this with $j>i$, we have $j-1>i$ and hence $d_{f_1} < d_{o_{m+1}}$.

If $o_{m+1}$ is irrelevant with respect to $o_m$, due to the second property of Theorem \ref{Prp1}, $d_{o_m} < d_{o_{m+1}}$ and $o_{m+1}$ cannot belong to a GOP with an index less than $j$. As a result, $d_{f_1} < d_{o_{m+1}}$ and $o_{m+1}$ belongs to a GOP with an index larger than $i$.

The second part can be similarly proven using mathematical induction.

\subsection{Proof of Lemma \ref{thm7} \label{AP5}}

We prove the first statement using mathematical induction.

For the basis step, consider $k=1$.  Since $o_1$ is scheduled after $f_1$ and they are irrelevant, due to the second property of Theorem \ref{Prp1}, we have $d_{f_1} < d_{o_1}$.

For the induction step, suppose the statement holds for $k=m$.  Consider the case $k=m+1$. We first note that none of $o_1,\ldots, o_k$ can be an I-frame, since otherwise $f_1$ would be scheduled after the I-frame. 
The inductive hypothesis says $d_{f_1} < d_{o_m}$. The frame $o_{m+1}$ either is a descendant of $o_m$ or is irrelevant with respect to $o_m$. If it is a descendant, then according to Lemma \ref{thm8}, $d_{f_1} < d_{o_{m+1}}$. If $o_{m+1}$ is irrelevant with respect to $o_m$, due to the second property of Theorem \ref{Prp1}, $d_{o_m} < d_{o_{m+1}}$. As a result, $d_{f_1} < d_{o_{m+1}}$.

The second statement can be proved similarly.

\subsection{Proof of Lemma \ref{lma0} \label{AP12}}

Assume towards a contradiction that the lemma does not hold for some $i$. Then it is easy to see that there must exist two frames, $f_1$ and $f_2$, in $GOP_i$ such that none of the frames scheduled between them belong to $GOP_i$.  Let the frames between $f_1$ and $f_2$ be $o_1,\ldots, o_k $.  In the SIO class, all of these frames are irrelevant with respect to $GOP_i$.  In particular, they are irrelevant with respect to $f_1$. As a result, Lemma \ref{lm1} suggests $o_k$ belongs to a $GOP_j$ such that $j>i$, which implies that $d_{f_2} < d_{o_k}$.  Noting that $o_k$ and $f_2$ are neighbors, this violates the second property of Theorem \ref{Prp1}.

\vspace*{-4mm}
\subsection{Proof of Lemma \ref{thm5} \label{AP6}}
Assume towards a contradiction that the lemma does not hold. Then it can be shown easily that frames $l'$, $l_1$, and $l_2$ exist such that  $l_1,l_2 \in T(l')$ and none of the frames scheduled between$l_1$ and $l_2$ belong to $T(l')$.

Without loss of generality, suppose $l_1$ is scheduled earlier than $l_2$.  Index the frames scheduled between $l_1$ and $l_2$ in the ascending order of transmission starting time to obtain $j_1,\ldots,j_{k'}$ for some integer $k'$. These frames are irrelevant with respect to $l'$ and therefore irrelevant with respect to $l_1$ and $l_2$. Hence, according to Lemma \ref{thm7},
\begin{equation}
d_{l_1} < d_{j_{m}} < d_{l_2}, 1\leq m \leq k'.
\label{eqn2}
\end{equation}
Now consider any specific $j_{m}$ among $j_1,\ldots, j_{k'}$. Lemma \ref{thm8} suggests that either $max\_dln(T(l')) < min\_dln(T(j_{m}))$ or $min\_dln(T(l')) > max\_dln(T(j_{m}))$ depending on the relative sizes of  $d_{j_{m}}$ and $d_{l'}$.  In the former case, since $l_1$ and $l_2$ belong to $T(l')$, both $d_{l_1}$ and $d_{l_2}$ are smaller than $d_{j_{m}}$, while in the latter case, similarly both $d_{l_1}$ and $d_{l_2}$ are greater than $d_{j_{m}}$. This contradicts \ref{eqn2}.

\section{Proof of Theorem \ref{thm10} \label{AP13}}

Let $\cal W$ be the canonical-form of a transmission sequence, with associated graph $G_{\cal W} \subset G$.  Since $G_{\cal W}$ can be obtained from $G$ by removing the nodes that are not scheduled in $\cal W$ and their descendants in the MBFS forest, the roots of trees in $G_{\cal W}$ are roots in $G$ as well.  Moreover, Lemma \ref{thm5} indicates that a node and its descendants in the MBFS forest are scheduled together.  This, together with the fact that the pre-order tree walk is executed on both $G$ and $G_{\cal W}$, implies that $\cal W$ can be obtained from the SIO universal sequence by removing subtrees, which corresponds to dropping frames.

\section{Proof of Lemmas \ref{lma2}-\ref{lma5}}
\subsection{Proof of Lemma \ref{lma2} \label{AP14}}

Towards a contradiction, suppose some frames $o_1,\ldots,o_k$ are scheduled between them. Since none of them  are in ${\cal N}_i$, they are all irrelevant with respect to $f_1$, so Lemma \ref{lm1} indicates $d_{f_1} < d_{o_k}$ and $o_k$  belongs to some $GOP_j$ such that $j>i$.  However, if $j=i+1$, this will violate the first property of Theorem \ref{Prp1} since $o_k$ has appeared prior to $I_{i+1}$ in the schedule; if $j>i+1$, then $d_{I_{i+1}}<d_{o_k}$, which violates the second property of Theorem \ref{Prp1}.

\subsection{Proof of Lemma \ref{lma3} \label{AP15}}

Towards a contradiction, suppose some frames $v_1,\ldots, v_k$ are scheduled between them. Since none of them are in ${\cal N}_i$, they are all irrelevant with respect to $f_2$, so Lemma \ref{lm1} indicates $d_{v_1} < d_{f_2}$ and $v_1$ belongs to some $GOP_j$ such that $j<i$. As a result, $v_1$ and $I_{i+1}$ are irrelevant and $d_{v_1} < d_{I_{i+1}}$. This violates the second property of Theorem \ref{Prp1}.

In addition, $f_2$ must be a descendant of $I_{i+1}$, since otherwise it will violate the second property of Theorem \ref{Prp1}.

\subsection{Proof of Lemma \ref{lma4} \label{AP16}}

Let the frames between $I_{i+1}$ and $f$ be $v_1,\ldots, v_k$. Two cases are considered: first, $I_{i+2}$ is among them, and second, it is not.

In the first case, Lemma $\ref{lma3}$ indicates $v_k=I_{i+2}$. Then, none of the frames $v_1,\ldots, v_{k-1}$ is an ancestor of $v_k$. Furthermore, the first property of Theorem \ref{Prp1} suggests that none of them can be a descendant of $v_k$ either. Hence, all frames $v_1,\ldots, v_{k-1}$ are irrelevant with respect to $v_k$, and Lemma \ref{lm1} indicates they all belong to GOPs with indices less than $i+2$. Since $v_1,\ldots, v_k$ do not belong to ${\cal N}_{i+1}$, they all belong to GOPs with indices less than $i+1$.  The same result can be obtained in the second case, with Lemma \ref{lm1} applied to frame $f$ instead of $v_k$.

Furthermore, in both cases, $v_1$ cannot be irrelevant with respect to $I_{i+1}$, since that will violate the second property of Theorem \ref{Prp1}. Then the first property of Theorem \ref{Prp1} indicates $v_1$ must be a descendant of $I_{i+1}$. As a result, $v_1 \in {\cal N}_i$ in both cases.  Now let $\xi=\max \{i : 1\leq i \leq k, v_i \in {\cal N}_i \}$. Lemma \ref{lma1} indicates that $v_1, \ldots, v_{\xi}$ all belong to ${\cal N}_i$.

In the first case, suppose $\xi< k-1$. Then the frames $v_{\xi+1}, \ldots, v_{k-1}$ belong to neither $GOP_i$ nor $GOP_{i+1}$. Therefore, they are all irrelevant with respect to $v_{\xi}$, and due to Lemma \ref{lm1}, they all belong to GOPs with indices higher than $i$. In addition, since none of them belong to ${\cal N}_{i+1}$, they all belong to GOPs with indices higher than $i+1$. But this contradicts the result above, so $\xi=k-1$. Similarly, in the second case, we can show that $\xi=k$.

\vspace*{-2mm}
\subsection{Proof of Lemma \ref{lma6} \label{AP11}}

Towards a contradiction, let $o_1,\ldots, o_k$ be the sequence of frames scheduled between $I_{i+1}$ and $\zeta_i$. These frames belong to ${\cal N}_i$ based on Lemma \ref{lma3}.  Frame $o_1$ cannot be in ${\cal N}_i-{\cal D}_i$, since otherwise the second property of Theorem \ref{Prp1} would be violated, due to $d_{o_1} < d_{I_{i+1}}$.  Hence,
$o_1 \in {\cal D}_i$, and $o_1$ is a critical node.  Then by the definition of $\zeta_i$, $d_{\zeta_i} <d_{o_1}$, and $\zeta_i$ and $o_1$ are irrelevant. Furthermore, by Theorem \ref{Prp1}, each frame $o_k$, $k>1$, is either the descendant of $o_{k-1}$ or irrelevant with respect to $o_{k-1}$, and $d_{o_{k-1}}<d_{o_k}$. Hence, the frames $o_2, \ldots, o_k$ are either descendants of $o_1$, or irrelevant with respect to $o_1$ but have a display deadline greater than $d_{o_1}$.  Frames in the latter case cannot be ancestors of $\zeta_i$, since otherwise according to Lemma \ref{thm8}, $d_{o_1}<d_{\zeta_i}$, which is a contradiction. Therefore, $o_2, \ldots, o_k$ are irrelevant with respect to $\zeta_i$.
In addition, based on Lemma \ref{thm8}, all of $o_1, \ldots, o_k$, especially $o_k$, have a deadline greater than $d_{\zeta_i}$. However, since $o_k$ and $\zeta_i$
are neighbors, this violates the second property of Theorem \ref{Prp1}.

\vspace*{-3mm}
\subsection{Proof of Lemma \ref{lma5} \label{AP8}}

For the first statement, assume towards a contradiction that the statement does not hold, i.e., there exists some frame in ${\cal N}_i-{\cal D}_i$, such that it is irrelevant with respect to $\zeta_i$ and has a deadline earlier than $d_{\zeta_i}$, and it is scheduled after $I_{i+1}$.  Let $x$ be the one of such frames that has the earliest scheduled transmission start time.  Then, by definition $d_x < d_{\zeta_i}$.  Furthermore, let $y$ be the frame scheduled immediately prior to $x$.  As explained previously, Lemmas \ref{lma1}, \ref{lma2}, \ref{lma3}, and \ref{lma4} jointly dictate that in the quasi-SIO canonical schedule all frames in ${\cal N}_i \cup \{I_{i+1}\}$ must be sent together.  Hence, either $y \in {\cal N}_i$ or $y = I_{i+1}$.   However, since $x\neq \zeta_i$, by Lemma \ref{lma6}, $y$ cannot be $I_{i+1}$.  Therefore, $y$ is scheduled between $I_{i+1}$ and $x$, and only the following two cases are left for $y$:

\begin{enumerate}
 \item $y \in {\cal N}_i-{\cal D}_i$

 In this case, since $y \notin {\cal D}_i$ clearly $y$ is not $\zeta_i$ nor a descendant of $\zeta_i$.  Furthermore, by the first property of Theorem \ref{Prp1} and Lemma \ref{lma6}, all ancestors of $\zeta_i$ other than $I_{i+1}$ must be scheduled ahead of $I_{i+1}$.  Therefore,  $y$ is not an ancestor of $\zeta_i$.  Hence, $y$ is irrelevant with respect to $\zeta_i$.  Furthermore, we have $d_{\zeta_i} <d_y$, since otherwise the existence of $y$ would contradict the definition of $x$.

 ~~This leads to two observations.  First, $d_x<d_y$, which implies that $x$ and $y$ cannot be irrelevant, due to the second property of Theorem \ref{Prp1}.  Second, Lemma \ref{thm8} suggests that $max\_dln(T(\zeta_i))<min\_dln(T(y))$.  This implies that $y$ cannot be an ancestor of $x$, since otherwise we would have $d_{\zeta_i} < d_x$, contradicting the definition of $x$.

 ~~Finally, the first property of Theorem \ref{Prp1} suggests that $y$ is not a descendant of $x$, since $y$ is schedule ahead of $x$.  Hence, this case is not possible.

 \item $y \in {\cal D}_i$

 We first note that $x$ cannot be a descendant of $y$ since $x \notin {\cal D}_i$, and $x$ cannot be an ancestor of $y$ due to the first property of Theorem \ref{Prp1}.  Hence, $x$ and $y$ are irrelevant.

 ~~Let $z$ be the critical frame such that $y$ belongs to $T(z)$ (with possibly $y=z$).  If $z =\zeta_i$, then since $x$ and $\zeta_i$ are irrelevant, Lemma \ref{thm8} indicates $max\_dln(T(x))<min\_dln(T(z))$, so we have $d_x < d_y$.  If $z \neq \zeta_i$, then $z$ and $\zeta_i$  are irrelevant and $d_{\zeta_i}<d_z$; therefore Lemma \ref{thm8} indicates $d_{\zeta_i}< d_y$.   Hence, in both cases, $d_x < d_y$.  This violates the second property of Theorem \ref{Prp1}.
\end{enumerate}

For the second statement, assume towards a contradiction that the statement does not hold, i.e., there exists some frame in ${\cal N}_i-{\cal D}_i$, such that it is irrelevant with respect to $\zeta_i$ and has a deadline later than $d_{\zeta_i}$, and it is scheduled before $I_{i+1}$.  Let $x$ be the one of such frames that has the latest scheduled transmission start time.  Then, by definition $d_{\zeta_i}< d_x$.  Let $y$ be the frame scheduled immediately after $x$.  As explained previously, Lemmas \ref{lma1}, \ref{lma2}, \ref{lma3}, and \ref{lma4} jointly dictate that in the quasi-SIO canonical schedule all frames in ${\cal N}_i \cup \{I_{i+1}\}$ must be sent together.  Hence, we have $y \in {\cal N}_i$ or $y = I_{i+1}$.  

\begin{enumerate}
 \item $y \in {\cal N}_i$
 
 In this case, since $y$ is scheduled before $I_{i+1}$, the first property of Theorem \ref{Prp1} suggests that $y \notin {\cal D}_i$.  This implies that $y$ cannot be $\zeta_i$ or a descendant of $\zeta_i$.
 
 ~~Now, suppose $y$ is an ancestor of $\zeta_i$. Then, $x$ cannot be an ancestor of $y$, since otherwise $x$ would not be irrelevant with $\zeta_i$.  However, $x$ cannot be a descendant of $y$ either, due to the first property of Theorem \ref{Prp1}.  Therefore, $x$ and $y$ are irrelevant.  Then, since $x$ and $y$ are scheduled next to each other, the second property of Theorem \ref{Prp1} suggests that $d_x < d_y$, which implies $max\_dln(T(x)) < min\_dln(T(y))$ by Lemma \ref{thm8}.  However, this also implies that $d_x < d_{\zeta_i}$, which contradicts the definition of $x$.  Hence, we conclude that $y$ cannot be an ancestor of $\zeta_i$.  

 ~~The above implies that $y$ and $\zeta_i$ are irrelevant.  Hence, we have $d_y < d_{\zeta_i}$, since otherwise the existence of $y$ would contradict the definition of $x$.  This leads to two observations.  First, $d_y < d_x$, which implies that $x$ and $y$ cannot be irrelevant, due to the second property of Theorem \ref{Prp1}.  Second, $y$ cannot be a descendant of $x$.  This is because $max\_dln(T(\zeta_i)) < min\_dln(T(x))$ by Lemma \ref{thm8}, so we would have $d_{\zeta_i} < d_y$ if $y$ were a descendant of $x$.
 
 ~~Furthermore, by the first property of Theorem \ref{Prp1}, $y$ cannot be an ancestor of $x$, since $y$ is schedule after $x$.  Hence, this case is not possible.

\item $y=I_{i+1}$

In this case, Lemma \ref{lma6} suggests that $y=I_{i+1}$ is followed immediately by $\zeta_i$.  Hence, we have the subsequence $x I_{i+1}\zeta_i$, with $d_{\zeta_i} < d_x$.  Furthermore, since $x\in {\cal N}_i-{\cal D}_i$, it is irrelevant with respect to $I_{i+1}$.  This violates the second property of Theorem \ref{Prp1}.

\end{enumerate}

\section{Proof of Theorem \ref{thm11} \label{AP17}}

The proof is similar to that of Theorem \ref{thm10}. We only need to additionally consider the case where an I-frame $I_{i+1}$ is not included in the transmission sequence.  We note that the removal of $I_{i+1}$ induces the removal of entire blocks ${\cal D}_{i}$ and ${\cal N}_{i+1}$, which does not alter the canonical form.

\vspace*{-2mm}

\bibliography{Papersfile}
\bibliographystyle{IEEEtran}

\end{document}